\pdfoutput=1
\documentclass[final,5p,times,twocolumn]{elsarticle}

\let\today\relax
\makeatletter
\def\ps@pprintTitle{%
    \let\@oddhead\@empty
    \let\@evenhead\@empty
    \def\@oddfoot{\footnotesize\itshape
         {} \hfill\today}%
    \let\@evenfoot\@oddfoot
    }
\makeatother

\usepackage[utf8]{inputenc} %unicode support
\usepackage[british]{babel}
\usepackage{multirow}
\usepackage[caption=false]{subfig}
\usepackage[hypcap=false]{caption}
\usepackage{graphicx}
\usepackage{amssymb}
\usepackage[cmex10]{amsmath}
\usepackage{graphicx}
\usepackage{mathrsfs}
\usepackage{verbatim}
\usepackage{epstopdf}
\usepackage{adjustbox}
\usepackage{color}
\usepackage{makecell}
\usepackage{bm}
\usepackage{hyperref}
\usepackage{enumitem}
\usepackage{times}
\usepackage{setspace}
\usepackage[figuresright]{rotating}
\usepackage{algorithm}
\usepackage{algorithmic}

\begin{document}

\begin{frontmatter}


\title{Adaptive Direction-Guided Structure Tensor Total Variation}



\author[rvt,focal]{Ezgi Demircan-Tureyen\corref{cor1}}
\ead{e.demircan@iku.edu.tr}

\author[focal]{Mustafa E. Kamasak}
\ead{kamasak@itu.edu.tr}

\cortext[cor1]{Corresponding author}

\address[rvt]{Dept. of Computer Engineering, Istanbul Kultur University, Istanbul 34156, Turkey}
\address[focal]{Dept. of Computer Engineering, Istanbul Technical University, Istanbul 34390, Turkey}

\begin{abstract}
Direction-guided structure tensor total variation (DSTV) is a recently proposed regularization term that aims at increasing the sensitivity of the structure tensor total variation (STV) to the changes towards a predetermined direction. Despite of the plausible results obtained on the uni-directional images, the DSTV model is not applicable to the multi-directional images of real-world. In this study, we build a two-stage framework that brings adaptivity to DSTV. We design an alternative to STV, which encodes the first-order information within a local neighborhood under the guidance of spatially varying directional descriptors (i.e., orientation and the dose of anisotropy). In order to estimate those descriptors, we propose an efficient preprocessor that captures the local geometry based on the structure tensor. Through the extensive experiments, we demonstrate how beneficial the involvement of the directional information in STV is, by comparing the proposed method with the state-of-the-art analysis-based denoising models, both in terms of restoration quality and computational efficiency.
\end{abstract}


\end{frontmatter}

%\linenumbers

\section{Introduction}
\label{sec:intro}
Restoration of an image from its noisy linear transformations is an active field of research in imaging science. This field involves a variety of linear inverse problems; such as denoising, deblurring, inpainting, and super-resolution. Among them, the denoising problem is extensively studied; not only due to having a wide spectrum of application fields, but also because it serves as a building block for the other, more complex problems. The fundamental drawback of denoising (and the others) is ill-posedness. The knowledge deduced from the observed image is often insufficient and requires to take some prior knowledge on the latent image into account. When such a problem is handled from a variational perspective, its reinterpretation in the form of an optimization problem involves a regularization term (i.e., regularizer), which is responsible for encoding that prior knowledge. 

A lot of effort has been dedicated to design good regularizers, which can characterize the unknown image well and in an efficiently solvable way. Rudin-Osher-Fatemi's total variation (TV) \cite{rudin1992nonlinear} has been considered to be a \textit{good} regularizer, due to its well-characterization of the images (as piecewise constant signals up to the edge regions), and its convex definition. Inevitably, TV's piecewise constancy (PWC) assumption does not always hold.  The real-world images may be piecewise smooth rather than constant, and when this is the case, TV yields artificial edges, known as staircase artifacts \cite{chan2000high}. As being the initiator of the notion of edge-preserving regularization, TV has given rise to a diverse range of TV inspired regularizers. Many of these regularizers extend TV to stretch its PWC assumption, by bring anisotropicity on the TV-term to better catch angled boundaries (e.g., \cite{esedoglu2004decomposition, chartrand2007exact, lou2015weighted}), and/or by incorporating the higher order information to promote piecewise smoothness (e.g., total generalized variation -- TGV \cite{bredies2010total}, and the other notable works \cite{chan2000high, lefkimmiatis2013hessian, papafitsoros2014combined}). 

There are also TV extensions that steer anisotropicity by a desired direction. %(e.g., \cite{steidl2009anisotropic}).******   
Directional total variation (DTV) \cite{bayram2012directional} is such a regularizer that aims at penalizing the local changes along a single dominant direction more than the others. For this purpose, it employs rotated and scaled gradients. Even though it performs well on the uni-directional images, DTV model fails to restore arbitrary, real-world images. It is applicable only when the image to be restored exhibits a single directional dominance. In \cite{zhang2013edge}, the authors suggested an edge-adaptive DTV (EADTV) to make DTV handle several dominant directions. They proposed a two-stage approach: (1) extracting the tangential directions of the edges, (2) feeding them into the EADTV model. However, the noise heavily affects the edges and mostly the wrong directions are penalized. A very recent work \cite{pang2020image} proposed an extended DTV model that follows a similar two-stage approach. Distinctively, to extract directions, they employed a more sophisticated orientation field estimation (OFE) method used for fingerprint identification in \cite{hong1998fingerprint}. They also made use of the mixed $\ell_p - \ell_2$ norm, where $p \in (0,1)$, while $p=1$ in DTV.

The direction-guided regularizers mentioned above requires a preprocessor that provides the directions to them. So, the erroneous outputs of the preprocessor mislead the succeeding inversion process. In \cite{grasmair2010anisotropic}, anisotropic TV filtering (ATV) was proposed as a combination of TV and anisotropic diffusion. It embodies the \textit{diffusion tensor} designed according to the \textit{structure tensor}. It intrinsically depends on the underlying image, rather than the observation. The downsides of ATV are its non-convexity and heuristic design. The objective function mostly has multiple local or global minimizers, and it may result in a suboptimal solution. The recent structure tensor total variation (STV) \cite{lefkimmiatis2015structure}, which forms the basis of this study in collaboration with DTV \cite{bayram2012directional}, proposed a convex objective by following the same idea behind ATV. It suggests to penalize the rooted eigenvalues of the structure tensor. In this way, it can capture the variations within a local neighborhood in contrast to TV and its localized extensions. In \cite{lefkimmiatis2015structure}, its superiority has been shown over vectorial TV (VTV) \cite{blomgren1998color}, which extends TV to handle vector-valued images, and TGV \cite{bredies2010total}.

Above mentioned regularizers are all local, except ATV and STV, which are classified as semi-local. There are also variational models that benefit from the non-local (NL) self similarity by adopting the idea of \cite{buades2011non} (e.g., NLTV \cite{gilboa2008nonlocal}, NLSTV \cite{lefkimmiatis2015nonlocal}). They usually perform better than their local or semi-local counterparts. Moreover, the regularizers mentioned here are all categorized as analysis-based priors, where the regularization directly runs on the image. Contrary to this, there are also synthesis-based priors that perform the regularization in a sparsifying transform domain (e.g., wavelet \cite{chambolle1998nonlinear, figueiredo2007majorization}, shearlets \cite{kutyniok2012shearlab}). Also there is a recent paper \cite{liu2018image} that proposes a semi-local regularizer along with its non-local version. Their prior depends on a new paradigm: It employs sparsity of the corners rather than the the edges. They showed that their method (Noble’s Corner Detector based Regularization -- NCDR) is superior to STV (and NLSTV for the non-local versions) in terms of restoration quality, apart from the fact that it is non-convex, works heuristically and slow.
  
%today's deep learning techniques (DnCNN \cite{zhang2017beyond}, FFDNet \cite{zhang2018ffdnet}, UNet and its non-local version UNLNet. Note that, our goal in this paper is to make a contribution to the analysis-based regularization literature.  

%Also, the dominance is characterized beforehand with two variables: for the strength of the dominance, and for the orientation of it. Thus, this information needs to be extracted from the observed data, before starting to the inversion process. 

\subsection{Contributions}
Our intention is proposing a variational denoising framework which produces physically satisfying and statistically meaningful results that are superior to the state-of-the-art analysis-based variational methods. In this respect, our contributions are listed below:
\begin{enumerate}
\item We design an adaptive direction-guided penalty term by extending STV.
\item We propose a preprocessing method that estimates the parameters required by our regularizer. 
\item We develop an efficient algorithm based on the proximal map of the proposed regularizer in order to solve the introduced convex optimization problem.
\item We assessed the performance of our framework by comparing the results with those obtained by TV, EADTV \cite{bayram2012directional, zhang2013edge}, STV \cite{lefkimmiatis2015structure}, and NCDR \cite{liu2018image}.
\end{enumerate}
Note that, we only focus on extending the semi-local STV to semi-local adaptive DSTV; thus, we won't consider designing a non-local counterpart of our regularizer, which can straightforwardly be accomplished by exploiting NLSTV \cite{lefkimmiatis2015nonlocal}. It would surely boosts the restoration quality.

%our goal in this paper is to make a contribution to the analysis-based regularization literature. Accordingly, \color{red}{mention deep learning, DnCNN etc.} \color{black} Also, 

\section{Problem Formulation}
We deal with the recovery of an underlying function $f: \Omega \mapsto \mathbb{R}^{C}$ from an observation $g:\Omega \mapsto \mathbb{R}^{C}$, where the forward corruption is modeled as: 
$g(x) = f(x) + \eta(x)$, where $\eta: \Omega \mapsto \mathbb{R}^{C}$ referring to the additive noise. Here, $C$ denotes the number of channels, which is more than one for the vector-valued images, and $x \in \Omega$, where the domain $\Omega \subseteq \mathbb{R}^{2}$ is a $2$-dimensional image space. The noise at each pixel is modeled as an \textit{i.i.d.} random variable that follows Gaussian distribution with $\eta(x) \sim \mathcal{N}(0,\sigma_{\eta}^2)$. 

The variational models combine a data fidelity term with a regularization term $R(f)$ in the form of an optimization problem. For the Gaussian denoising, the fidelity term is nothing but the least squares fitting term to stick by the observed data, thus, one has the objective function of the form:
\begin{equation}
\label{eq:energy} 
E(f) = \frac{1}{2} \Vert g - f\Vert_2^2  + \tau R(f)
\end{equation}
to be minimized with respect to $f$. Here, $\tau$ is the regularization parameter that is used to tune the contribution of $R(f)$ to the cost. Note that, there are studies in the literature that makes $\tau$ adaptive to give rise to spatially dependent regularization (e.g., \cite{gilboa2006variational}, \cite{grasmair2009locally}, \cite{dong2009multi}).

\subsection{Notation}
Throughout the paper, both of the scalar-valued grayscale images ($C=1$) and the vector-valued RGB images ($C=3$) will be our concern. We will only focus on the discrete domain, where $f$ is a discrete-space function, thus we switch the notation from $f(x)$ to $f[i]$, for $i \in \lbrace 1, 2, \cdots, N \rbrace$ referring to the index of a certain pixel, where $N$ is the number of pixels in a single channel. Note that, we treat the images as being lexicographically stacked into vectors ($f \in  \mathbb{R}^{NC}$). 

$\Vert X \Vert_{p,q}$ stands for either the mixed $\ell_p - \ell_q$ norm defined as $(\sum_{i=1}^{N} \Vert X[i] \Vert_q^p)^{1/p}$, or the mixed $\ell_p - \mathcal{S}_q$ norm when X is a matrix. Here $\mathcal{S}_q$ refers to the Schatten norm of order $q$\footnote{Schatten norm of a matrix $\textbf{X}$ is $\ell_q$ norm of the vector $\boldsymbol{\sigma} (\textbf{X})$, where $\sigma$ returns the singular values, i.e., $\Vert \textbf{X} \Vert_{\mathcal{S}_q} = \Vert
\boldsymbol{\sigma} (\textbf{X}) \Vert_q$ \cite{bhatia2013matrix}}. The symbols $\odot$ and $^\circ$ are used to denote entrywise product and entrywise power, respectively; while $\langle \cdot \rangle$ refers to the inner product. Furthermore, the vectors/matrices are denoted by using lowercase/uppercase bold letters thereafter. 

\section{Background}
\label{sec:background}
The analysis-based regularizers can generically be defined as follows:
\begin{equation}
\label{eq:analysis-based} 
R(\textbf{f}) = \Vert  \phi((\Gamma \textbf{f})[i]) \Vert_{p}
\end{equation}
TV seminorm is nothing but the $\ell_1$ norm of the gradient magnitude, i.e., according to Eq. \eqref{eq:analysis-based}, $\Gamma=\nabla$, $\phi = \Vert \cdot \Vert_2$, and $p=1$, where ${\nabla}$ denotes the gradient operator.  As mentioned in Sect.~\ref{sec:intro}, many alternatives to this native isotropic TV have been proposed, e.g., $\phi = \Vert \cdot \Vert_1$ in \cite{esedoglu2004decomposition}, $\phi = \Vert \cdot \Vert_q$ and $q \in (0,1)$ in \cite{chartrand2007exact}, $\phi = w_1 \Vert \cdot \Vert_1 + w_2 \Vert \cdot \Vert_2$ in \cite{lou2015weighted}, where $\Gamma={\nabla}$ and $p=1$ were kept for all.  

%\begin{equation}
%\label{eq:analysis-based} 
%TV(f) = \Vert (\nabla f)[i] \Vert_{1,2}
%\end{equation}

\subsection{Directional Total Variation (DTV)}
\label{subsec:DTV}
DTV \cite{bayram2012directional} is a local analysis-based regularizer that simply changes the TV's rotation invariant ${\nabla}$ in Eq. \eqref{eq:analysis-based} with ${\Gamma}={\boldsymbol{\Lambda}}_{\alpha} {\textbf{R}}_{-\theta}{\nabla}$, where ${{\textbf{R}}}_{\theta} = \left[ \begin{array}{cc}
\cos\theta & -\sin\theta \\
\sin\theta & \cos\theta
\end{array} \right]$ and ${\boldsymbol{\Lambda}}_\alpha = \left[ \begin{array}{cc}
\alpha & 0 \\
0 & 1
\end{array} \right]$ with $\theta \in [0, \pi)$ corresponding to the dominant direction $(\cos\theta, \sin\theta)$, and $\alpha$ that is used to tune the dose of the penalization. In this way, it penalizes  the magnitudes of the scaled and rotated gradients, towards a predetermined direction. As mentioned earlier, DTV can only be applied to the uni-directional images. EADTV \cite{zhang2013edge} is its adaptive extension that suggests to use spatially varying angles $\boldsymbol{\theta}[i]$.  
Even it is trivial to estimate a scalar $\theta$ from an observed uni-directional image, when $\boldsymbol{\theta}$ represents an \textit{orientation field}, it becomes a challenging problem. EADTV simply computes each $\boldsymbol{\theta}[i]$ such that,
\begin{equation}
\label{eq:edgeadaptivity} 
(\cos\boldsymbol{\theta}[i], \sin\boldsymbol{\theta}[i]) \perp ({\nabla} \textbf{g}_{\sigma})[i]
\end{equation}
where the subscript ${\sigma}$ indicates that the image $\textbf{g}$ is pre-smoothed by a Gaussian filter of standard deviation $\sigma$. However, the gradient is not a convenient descriptor under extensive amount of noise. The estimation of the gradient is mostly misguided, and since DTV is sensitive along the angles $\boldsymbol{\theta}[i]$, the structure is destroyed by the incorrect estimates. Moreover, rippled values of the angles cause unstable output.

By recalling Eq. \eqref{eq:analysis-based}, the recent paper \cite{pang2020image} plugs $\Gamma = {{\Lambda}}_{\alpha}(\textbf{g}) {{R}}_{-\theta(\textbf{g})} \nabla $, $\phi = \Vert \cdot \Vert_2$, and $p \in (0,1)$\ where ${\Lambda}_\alpha$ and ${\theta}$ are both functions of $\textbf{g}$. Except the preference of the outer norm, this definition differs from the EADTV in the sense that it better recognizes the local edge orientations due to the more complex OFE method that it employs, and uses an adaptive stretching matrix weighted according to the first-order local edge information. 

\subsection{Structure Tensor Total Variation (STV)}
The \textit{structure tensor} of an image $\textbf{f}$ at a point $i$ is a $2 \times 2$ symmetric positive semi-definite (PSD) matrix of the form:
\begin{equation}
\label{eq:structure_tensor}
({S}_{K}\textbf{f})[i] = \textbf{K}_{\hat{\sigma}} * ((J \textbf{f})[i] (J \textbf{f})[i]^T)
\end{equation}
where $\textbf{K}_{\hat{\sigma}}$ is a Gaussian kernel of standard deviation ${\hat{\sigma}}$ centered at the $i$-th pixel, and $J$ denotes the Jacobian operator that extends the gradient for vector-valued images, i.e.,
\begin{equation}
\label{eq:Jacobian}
(J\textbf{f})[i] = \big[(\nabla \textbf{f}^{1})[i] \ (\nabla \textbf{f}^{2})[i] \ \cdots \ (\nabla \textbf{f}^{C})[i]\big]
\end{equation}
The superscripted $\textbf{f}$'s are denoting the channels, and for scalar-valued images, $J(\cdot) = \nabla(\cdot)$. This semi-local descriptor can summarize all the gradients within the patch supported by $\textbf{K}_{\hat{\sigma}}$, thus provides a better way of characterizing the variations, when compared to the local differential operators. The eigenvectors of it correspond to the directions of maximum and minimum vectorial variations, and the rooted eigenvalues serve as a measure of variations towards the respective directions. At this point, the STV prior \cite{lefkimmiatis2015structure} presents an innovatory way of describing the total amount of variations. It replaces the TV's $\nabla$ in Eq. \eqref{eq:analysis-based} with $\Gamma = \sqrt{\boldsymbol{\lambda}(\cdot)}$, where $({\boldsymbol{\lambda}\textbf{f})[i]}$ is a 2D vector composed of the maximum $\lambda^+(\cdot)$ and the minimum $\lambda^-(\cdot)$ eigenvalues of the structure tensor at the point $i$, i.e., ${(\boldsymbol{\lambda}\textbf{f})[i]} = [{\lambda^+(({S}_{K}\textbf{f})[i])}, {\lambda^-(({S}_{K}\textbf{f})[i])} ]^T$. STV also considers a more general case by using $\phi = \Vert \cdot \Vert_q$ ($q>1$).
%\begin{equation}
%\label{eq:stv1}
%STV(f) = \sum_{i=1}^N \Vert \sqrt{\lambda[i]} \Vert_p 
%\end{equation}
Since this design of STV is nonlinear and Eq. \eqref{eq:structure_tensor} involves a convolution kernel, the authors of \cite{lefkimmiatis2015structure} reformulated $({S}_K\textbf{f})[i]$ in terms of another operator that they named as \textit{patch-based Jacobian} $J_K: \mathbb{R}^{N \times C} \mapsto \mathbb{R}^{N \times LC \times 2}$, which embodies the convolution kernel of size $L$, and enables to express the STV functional such that it can be decomposed into linear functionals. $(J_K \textbf{f})[i]$ is defined as: 
\begin{equation}
\label{eq:patchbasedJacobian}
(J_K \textbf{f})[i] = \big[(\tilde{\nabla} \textbf{f}^{\textbf{1}})[i], \ (\tilde{\nabla} \textbf{f}^{\textbf{2}})[i] \ \cdots \ (\tilde{\nabla} \textbf{f}^{\textbf{C}})[i]\big]^T
\end{equation}
where $(\tilde{\nabla} \textbf{f}^{\textbf{c}})[i] = \big[(\mathcal{T}_{1}\nabla \textbf{f}^\textbf{c})[i], (\mathcal{T}_{2}\nabla \textbf{f}^\textbf{c})[i], \cdots, (\mathcal{T}_{L}\nabla \textbf{f}^\textbf{c})[i]\big]$. Each $l$-th entity applies shifting and weighting on the gradient as $(\mathcal{T}_{l} \nabla \textbf{f}^\textbf{c})[i] =  \sqrt{K_\sigma[p_l]}(\nabla \textbf{f}^\textbf{c})[x_i-p_l]$, where $x_i$ denotes the actual 2D coordinates of the $i$-th pixel, and for $L = (2L_K+1)^2$, $p_l \in \lbrace-L_K, \cdots,L_K \rbrace^2$ is the shift amount. 
%$\mathcal{P}$-neighborhood ($\mathcal{P} = \lbrace-L_K, \cdots, L_K \rbrace^2$) of $x_i$. 
That way, the structure tensor in terms of the patch-based Jacobian is:
\begin{equation}
\label{eq:structure_tensor_v2}
({S}_{K}\textbf{f})[i] = (J_K \textbf{f})[i] (J_K)\textbf{f})[i]^T
\end{equation}
Now, the rooted eigenvalues of $({S}_{K}\textbf{f})[i]$ coincide with the singular values of $(J_K \textbf{f})[i]$, and by employing $\mathcal{S}_q$ matrix norm, STV is redefined as: 
\begin{equation}
\label{eq:stv} 
STV(\textbf{f}) = \sum_{i = 1}^N \Vert (J_K\textbf{f})[i] \Vert_{\mathcal{S}_q}
\end{equation}
This redefinition allows one to find an efficient numerical solution to STV regularized problems, by using convex optimization tools.

%This idea was influenced by the directional TV (DTV) \cite{bayram2012directional}, which penalizes the magnitudes of the scaled and rotated gradients, towards a predetermined direction, characterized by an angle $\theta$. Both of the DTV and DSTV necessitate uni-directional images, which makes them impractical for the real world applications. In this paper, we aim at extending the DSTV, so that it can be applied to the images with arbitrary directions. Our approach treats such images as a collection of different-sized uni-directional patches. We contribute by proposing a preprocessor method that extracts  $\theta[i]$'s, varying from patch to patch, from the noisy image, and by extending DSTV functional to accept them. Our method utilizes the histogram of oriented gradients (HOG), which is a well-known feature descriptor, and the quadtrees to characterize these intrinsic directionalities. 

\section{Proposed Method}
%\subsection{Relation between DTV and STV}
Let us first point out the relation between the DTV and STV measures. By expanding the DTV measure, one can draw the following conclusion:
\begin{equation}
\label{eq:dtv_as_directionalDrv} 
DTV (\textbf{f}) \triangleq \Vert {\boldsymbol{\Lambda}}_{\alpha} {\textbf{R}}_{-\theta}{\nabla} \textbf{f} \Vert_{1,2} = \Bigg\Vert \left[ \begin{array}{c}
\alpha D_\theta \textbf{f} \\
D_{\theta^\bot} \textbf{f}
\end{array} \right] \Bigg\Vert_{1,2}
\end{equation}
where $D_\theta$ denotes the \textit{directional derivative} of $\textbf{f}$ in the direction characterized by $\theta$, and $\theta^\bot = \theta \pm \pi/2$. Eq. \eqref{eq:dtv_as_directionalDrv} shows that DTV actually works by increasing the dose of penalization (with the weight $\alpha$) in the direction that it presumes the variation is minimum. The penalty dose in the direction of maximum variation, which is orthogonal to that of the minimum, is only determined by the regularization parameter $\tau$. This idea of penalizing the maximum and minimum directional variations coincides with the STV. However, instead of making assumptions, STV codifies the maximum and the minimum directional variations with the maximum and the minimum eigenvectors and the square roots of the corresponding eigenvalues ($\sqrt{{\lambda}^+}$ and $\sqrt{{\lambda}^-}$) of the structure tensor. In other words, it deduces the directional variation from a summary of all derivatives within a local neighborhood.
\begin{equation}
\label{eq:stv_as_directionalDrv} 
STV (\textbf{f}) \triangleq \Bigg\Vert \left[ \begin{array}{c}
\sqrt{\lambda^+ (S_K\textbf{f})} \\
\sqrt{\lambda^- (S_K\textbf{f})}
\end{array} \right] \Bigg\Vert_{1,q}
\end{equation}
From this point of view, the question that motivated us to design a direction-guided STV was: Can STV more accurately deduce the directional variation if the summarized derivatives are directional in a pre-determined direction?

\subsection{Direction-Guided STV (DSTV)}
While it has been shown that the STV produces the state-of-the-art results among the other analysis-based local regularizers; under excessive amount of noise, it may not distinguish the edges and may smooth out them. At this point, the DSTV \cite{demircan2019direction} comes into play. It aims at incorporating the prior knowledge on the underlying image's local geometry into STV.
Under the inspiration of DTV, DSTV employs directional derivatives by applying the operator $\boldsymbol{\Pi}_{(\alpha, \theta)}$ that can act on the gradient at each image point as $(\boldsymbol{\Pi}_{(\alpha, \theta)} \nabla \boldsymbol{f}^{\boldsymbol{c}})[i] = \boldsymbol{\Lambda}_{\alpha} \textbf{R}_{-\theta} (\nabla \boldsymbol{f}^{\boldsymbol{c}}) [i]$, while gathering the neighboring gradients to constitute the patch-based Jacobian as:
\small
\begin{equation}
\label{eq:directionalPatchJacob} 
\begin{split}
( J^{(\alpha, \theta)}_{K}&\boldsymbol{f})[i] = \\&\left[ \begin{array}{cccc}
(\mathcal{T}_{1} \boldsymbol{\Pi}_{(\alpha, \theta)} \nabla \boldsymbol{f}^{\boldsymbol{1}})[i] & (\mathcal{T}_{2} \boldsymbol{\Pi}_{(\alpha, \theta)} \nabla \boldsymbol{f}^{\boldsymbol{1}})[i] & \cdots & (\mathcal{T}_{L}\boldsymbol{\Pi}_{(\alpha, \theta)} \nabla \boldsymbol{f}^{\boldsymbol{1}})[i] \\
(\mathcal{T}_{1} \boldsymbol{\Pi}_{(\alpha, \theta)} \nabla \boldsymbol{f}^{\boldsymbol{2}})[i] & (\mathcal{T}_{2}\boldsymbol{\Pi}_{(\alpha, \theta)} \nabla \boldsymbol{f}^{\boldsymbol{2}})[i] & \cdots & (\mathcal{T}_{L} \boldsymbol{\Pi}_{(\alpha, \theta)} \nabla \boldsymbol{f}^{\boldsymbol{2}})[i]\\
\vdots & \vdots & \vdots & \vdots\\
(\mathcal{T}_{1}\boldsymbol{\Pi}_{(\alpha, \theta)} \nabla \boldsymbol{f}^{\boldsymbol{c}})[i] & (\mathcal{T}_{2}\boldsymbol{\Pi}_{(\alpha, \theta)} \nabla \boldsymbol{f}^{\boldsymbol{c}})[i] & \cdots & (\mathcal{T}_{L}\boldsymbol{\Pi}_{(\alpha, \theta)} \nabla \boldsymbol{f}^{\boldsymbol{c}})[i]
\end{array} \right]
\end{split}
\end{equation}
$J^{(\alpha, \theta)}_K$ is named as \textit{directional patch-based Jacobian} \cite{demircan2019direction}, and involved by the DSTV functional, i.e.,
\begin{equation}
\label{eq:dstv1} 
DSTV(\textbf{f}) = \sum_{i = 1}^N \Vert (J^{(\alpha, \theta)}_K \textbf{f})[i] \Vert_{\mathcal{S}_q}
\end{equation}
\normalsize

By means of Eq. \eqref{eq:dstv1}, one has the chance of leading the STV machinery in favor of a predetermined direction. It has been experimentally shown on uni-directional images in \cite{demircan2019direction} that this incorporation produces prominently better results. Also, since the convexity is preserved in DSTV, the same convex optimization tools used to solve STV can be applied to DSTV based problems. %For the optimization algorithm that evaluates the proximal map of DSTV, the reader is referred to \cite{demircan2019direction}. 

\subsection{Adaptive DSTV (ADSTV)}
In natural images, the dominant directions vary in different regions, which makes DSTV inapplicable. In order to handle arbitrary natural images, we employ the same way that EADTV employed when bringing adaptivity to DTV. We consider $\theta$ as a spatially varying field of orientation, i.e., $\boldsymbol{\theta}: \Omega \mapsto [0, \pi)^C$, and this changes the rotation matrix to ${R}_{\boldsymbol{-\theta}[i]}$.  % = \left[ \begin{array}{cc}\text{cos}\theta[i] & \text{-sin}\theta[i] \\\text{sin}\theta[i] & \text{cos}\theta[i] \end{array} \right]$. 

Furthermore, in natural images, the anisotropic structures are not homogeneously distributed. Also among the anisotropic structures, their degrees of anisotropy spatially change. DSTV's stretching matrix $\Lambda_\alpha$ uses a constant factor $\alpha$ for the entire image. Just as done to the rotation parameter, one may employ a field of spatially changing stretching factors, so that the dose of penalization at a point can be tuned. However, since the regularization parameter $\tau$ is a constant scalar, with varying stretching factor the total amount of regularization changes from region to region. The fittest value of $\tau$ for a directional region that requires larger values of $\boldsymbol{\alpha}[i]$ may produce poor results in the regions with isotropic structures, where the corresponding values of $\boldsymbol{\alpha}[i]$ are small. To compensate it, we suggest to use a scaling matrix $\tilde{\boldsymbol{\Lambda}}_{(\alpha^+,\boldsymbol{\alpha}^\textbf{-}[i])}$ with two parameters, one of which is coming from a spatially varying field of $\boldsymbol{\alpha}^\textbf{-}:\Omega \mapsto [1,\alpha^+]^C$, i.e., 
\begin{equation}
\label{eq:Lambda} 
\tilde{\boldsymbol{\Lambda}}_{(\alpha^+,\boldsymbol{\alpha}^\textbf{-}[i])} = \left[ \begin{array}{cc}
\alpha^+ & 0 \\
0 & \boldsymbol{\alpha}^\textbf{-}[i]
\end{array} \right]
\end{equation}
While at a point $i$ within a highly directional region, $\boldsymbol{\alpha}^\textbf{-}[i]=1$; in a flat region, it switches to isotropic behavior: $\boldsymbol{\alpha}^\textbf{-}[i]=\alpha^+$. When it comes to the values between these two end points, as they get closer to one, the sensitivity to the changes towards $\boldsymbol{\theta}[i]$ increases. 

As a consequence, the directional patch-based Jacobian operator given in Eq. \eqref{eq:directionalPatchJacob} is extended to its adaptive version (which will be denoted as $\tilde{J}^{(\boldsymbol{\alpha},\boldsymbol{\theta})}_K$) that employs $(\tilde{\Pi}_{(\boldsymbol{\alpha},\boldsymbol{\theta})}\nabla \textbf{f}^c)[i] = \tilde{\boldsymbol{\Lambda}}_{(\alpha^+,\boldsymbol{\alpha}^\textbf{-}[i])} \textbf{R}_{\boldsymbol{-\theta}[i]}(\nabla \textbf{f}^c)[i]$ to perform directional derivation. Therefore, our ADSTV regularizer is defined as follows:
\begin{equation}
\label{eq:adstv1} 
ADSTV(\textbf{f}) \triangleq \sum_{i = 1}^N \Vert (\tilde{J}^{(\boldsymbol{\alpha},\boldsymbol{\theta})}_K \textbf{f})[i] \Vert_{\mathcal{S}_q} = \Vert \tilde{J}^{(\boldsymbol{\alpha},\boldsymbol{\theta})}_K \textbf{f} \Vert_{1,q} 
\end{equation}

\subsection{Parameter Estimation}
\label{DPE}
Estimation of the parameters $\boldsymbol{\alpha}^-$ and $\boldsymbol{\theta}$ is not a trivial task. In a noisy image, line-like structures are often interrupted by the noise, which results in deviations from the correct direction. When the conventional smoothing filters are applied to suppress noise, the same line-like structures get thicker, and the edges are dislocated. The reconnection of the interrupted lines is concerned in many studies, such as the ones that employ the morphological operators for the curvilinear structures in biomedical images \cite{merveille20172d, merveille2017curvilinear}, or for the ridges in fingerprint images \cite{oliveira2008multiscale}. Their principle goal is detection rather than restoration. %The main disadvantage of those methods is their high computational cost. 
In the proposed method, we focus finding a layout for the orientations rather than a precise field of them. By means of that, the preprocessor does not put much computational overhead. 

In our preprocessor, we employ eigenvectors and eigenvalues of the structure tensor to extract the knowledge about the orientations and the dose of anisotropy, in semi-local fashion, in contrast to EADTV. We also employ isotropic TV regularization to circumvent the problem of interrupted lines and to ignore the insignificant deviations. Below, we explain our parameter estimation procedure, which we will refer to as directional parameter estimator (DPE), step by step in detail. Note that our preprocessor employs Sobel operator as the discretization scheme used for the gradient. Also note that, our parameter estimator uses the luminance information of the input image ($\textbf{g}_L$), therefore for the vector-valued images, each channel of the $\boldsymbol{\theta}$ will be a copy of each other, i.e., $\boldsymbol{\theta}^1=\boldsymbol{\theta}^2=\cdots=\boldsymbol{\theta}^C$). The same applies to $\boldsymbol{\alpha}^-$. 

%The technique deals with the problem of completion of interrupted lines and enhancement of flow-like features in images. The completion of line-like structures is also a major concern in diffusion tensor magnetic resonance imaging (DT-MRI). 

As a measure of anisotropy, we equip a field of \textit{coherence} measures $\textbf{c}: \Omega \mapsto [0,1]$ computed at each image point as follows:
\begin{equation}
\label{eq:coherence}
\textbf{c}_{\sigma}[j] =  \frac{\lambda_{\sigma}^+[j] - \lambda_{\sigma}^-[j]}{\lambda_{\sigma}^+[j]}
\end{equation}
where $\lambda^\pm_{\sigma}[j] = \lambda^\pm(({S}_K {\textbf{g}_\textbf{L}}_\sigma)[j])$ and $j = \lbrace 1, 2, \cdots, N\rbrace$. Note that in this section, the variables subscripted by $\sigma$ denote the pre-smoothed version of $\textbf{g}_\textbf{L}$ not only with the Gaussian kernel of standard deviation $\sigma$, but also with the support $\sigma^2 \times \sigma^2$. Moreover, the kernel $K_{\bar{\sigma}}$ embodied by the structure tensor that we employ here operates $\bar{\sigma}$ larger than $\hat{\sigma}$, again with the support $\bar{\sigma}^2 \times \bar{\sigma}^2$. Eq. \eqref{eq:coherence} is nothing but a measure of the contrast between the directions of the highest and lowest fluctuations. For a pure local orientation ($\lambda_{\sigma}^+ > 0, \lambda_{\sigma}^- = 0$) the coherence is one, while for an isotropic structure ($\lambda_{\sigma}^+ = \lambda_{\sigma}^- > 0$) it is zero. After computing $\textbf{c}_\sigma$, by treating it as a corrupted image, we apply TV-based regularization. This way, the fluctuating intensities are suppressed, thus the uncertain responses are refused. The ROF model \cite{rudin1992nonlinear} that serves our purpose is given below:
\begin{equation}
\label{eq:coherenceTV}
\hat{\boldsymbol{\kappa}}_\sigma =  \underset{\boldsymbol{\kappa}_\sigma \in [0,1]^N}{\operatorname{argmin}} \ \Vert \textbf{c}_\sigma - \boldsymbol{\kappa}_\sigma \Vert_2^2 + TV(\boldsymbol{\kappa}_\sigma)
\end{equation}

To be able to detect all desired structures, the response of TV regularized coherence $\hat{\boldsymbol{\kappa}}_\sigma$ is analyzed at different scales (i.e., $\sigma_k^2 = 2k-1$, where $k = \lbrace 1,2,\cdots,K\rbrace$). Since as the scale parameter $\sigma_k$ gets larger, the false positive edge responses come into being; our method follows the iterative procedure shown below by giving more credits to the responses obtained at the fine-scales:
%For the noise levels less than $0.2$, we suggest using $K=2$, while $K=3$ for the higher ones. By relying on the response coming from the lower scales more than the higher ones, we design the following iterative procedure. For the initialization $\hat{\boldsymbol{\kappa}}^{(0)} = \hat{\boldsymbol{\kappa}}_{\sigma_{1}}$:  
%\small
\begin{equation}
\label{eq:scaleSpaceCoherence} 
{\boldsymbol{\kappa}}^{(k)}[j] =
    \begin{cases}
    {\boldsymbol{\kappa}}^{(k-1)}[j], & \text{if}\  \hat{\boldsymbol{\kappa}}_{\sigma_{k}}[j] \leq {\boldsymbol{\kappa}}^{(k-1)}[j]\\
    \frac{1}{2}({{\boldsymbol{\kappa}}^{(k-1)}[j]+\hat{\boldsymbol{\kappa}}_{\sigma_{k}}[j]}), & \text{otherwise}
    \end{cases}
\end{equation}
%\normalsize 
We suggest using $K=2$ for the noise levels less than $0.2$, while $K=3$ for the higher ones. If the image at hand is thought to be either highly directional or highly nondirectional, our algorithm further improves the coherence values. It proceeds by enhancing the values higher than the mean in a directional image, while diminishing the values less than the mean in a nondirectional one. This procedure is also executed at each scale, i.e.,
\begin{equation}
\label{eq:scaleSpaceCoherence2} 
{\boldsymbol{\phi}}^{(k)}[j] = \begin{cases} \mathcal{V}({\boldsymbol{\kappa}}^{(k)}[j],1), & \text{if}\  \gamma_1^{(k)} > 1\\ \mathcal{V}(\sqrt[\leftroot{-2}\uproot{2}4]{{\boldsymbol{\kappa}}^{(k)}[j]},-1), & \text{else if} \ \gamma_1^{(k)} < -1 \\ {\boldsymbol{\kappa}}^{(k)}[j], & \text{otherwise}\end{cases}
\end{equation}
Here the function $
\mathcal{V}(X,s) = \begin{cases} X^2, & \text{if}\   sX < s \mu^{(k)}\\ 
X, & \text{otherwise} \end{cases}$, where $\mu^{(k)}$ is the mean of the TV regularized coherence field, i.e., $\mu^{(k)} = E[\hat{\boldsymbol{\kappa}}^{(k)}]$, and the variable
$\gamma_1^{(k)}$ denotes the skewness of the TV regularized coherence histogram at the scale $\sigma_k$, i.e., $\gamma_1^{(k)} = Skew[{\boldsymbol{\kappa}}^{(k)}]$. If the skewness is greater than 1, the distribution of the coherence values is considered to be highly skewed to the right, which implies that the nondirectional regions predominate the directional ones. If it is less than -1, the distribution is highly left skewed and it can be inferred that the image is highly directional. Note that in our implementation, we used $skewness (\cdot)$ function of Matlab. 

Finally, our algorithm obtains the entries of $\boldsymbol{\alpha}^-$, by inverse scaling the entries of ${\boldsymbol{\phi}} \in [0,1]^N$ onto the range $[1, \alpha^+]$, so that the regions with higher response to the above mentioned coherence estimation procedure will be regularized by using smaller $\boldsymbol{\alpha}^-[j]$. 
\begin{equation}
\label{eq:alphaFinal}
\boldsymbol{\alpha}^-[j] = \frac{\alpha^+ - 1}{\text{max}({\boldsymbol{\phi}})-\text{min}({\boldsymbol{\phi}})} (\text{max}({\boldsymbol{\phi}}) - {\boldsymbol{\phi}}[j]) + 1
\end{equation}

When it comes to the estimation of $\boldsymbol{\theta}$, we first consider the eigendirections of the structure tensor, computed at a scale with the highest response to Eq. \eqref{eq:scaleSpaceCoherence}, i.e.,
\begin{equation}
\label{eq:theta1}
\boldsymbol{\vartheta}[j] = \angle \ \textbf{v}^-_{\hat{\sigma}_k}[j], \quad \hat{\sigma}_k = \underset{\sigma_1 < \cdots < \sigma_K}{\operatorname{argmax}}{\hat{\boldsymbol{\kappa}}_{\sigma_k}[j]}
\end{equation}
where $\textbf{v}^-_{\hat{\sigma}_k}[j]$ is the eigenvector corresponding to the smallest eigenvalue $\lambda^+_{\hat{\sigma}_k}[j]$. Next, as was done to the coherence, the estimated angles are TV regularized as follows: 
\begin{equation}
\label{eq:thetaTV}
\hat{\boldsymbol{\theta}} =  \underset{\boldsymbol{\theta} \in [0,\pi)^N}{\operatorname{argmin}} \ \frac{1}{2} \Vert {\boldsymbol{\vartheta}} - {\boldsymbol{\theta}} \Vert_2^2 + \tau TV(\boldsymbol{\theta})
\end{equation}
where we set $\tau = 0.02$ for all our experiments. Setting $\tau$ to a larger value may cause deviations from the correct directions due to the contrast loss.
\begin{comment}
\begin{equation}
\label{eq:theta}
\boldsymbol{\vartheta}^{(k)}[j] =
    \begin{cases}
    \angle \ \textbf{v}^+_{\sigma_k}[j], & \text{if}\  (\hat{\boldsymbol{\kappa}}_{\sigma_k}[j] >  \hat{\boldsymbol{\kappa}}^{(k-1)}[j]) \\ %\land (\boldsymbol{\vartheta}^{(k-1)}[j] = \text{NaN})\\
\boldsymbol{\vartheta}^{(k-1)}[j], & \text{otherwise}
    \end{cases}
\end{equation}
\end{comment}
 \begin{figure*}[t!]
 \includegraphics[width=\textwidth]{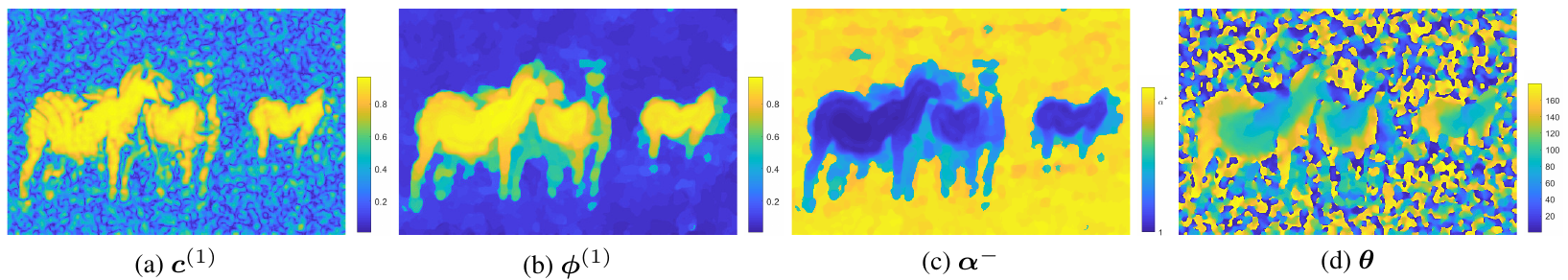}
  \caption{Exemplary outputs of the DPE procedure.
The colormaps respectively show (a) the coherence field of unsmoothed $\textbf{g}_L$, (b) its TV regularized version, (c) the final weights, and (d) the final directions for the noise level $\sigma_\eta = 0.15$.}
 \label{fig:DPE}       
    \end{figure*}
    
 \begin{figure}[t!]
 \includegraphics[width=8.8cm]{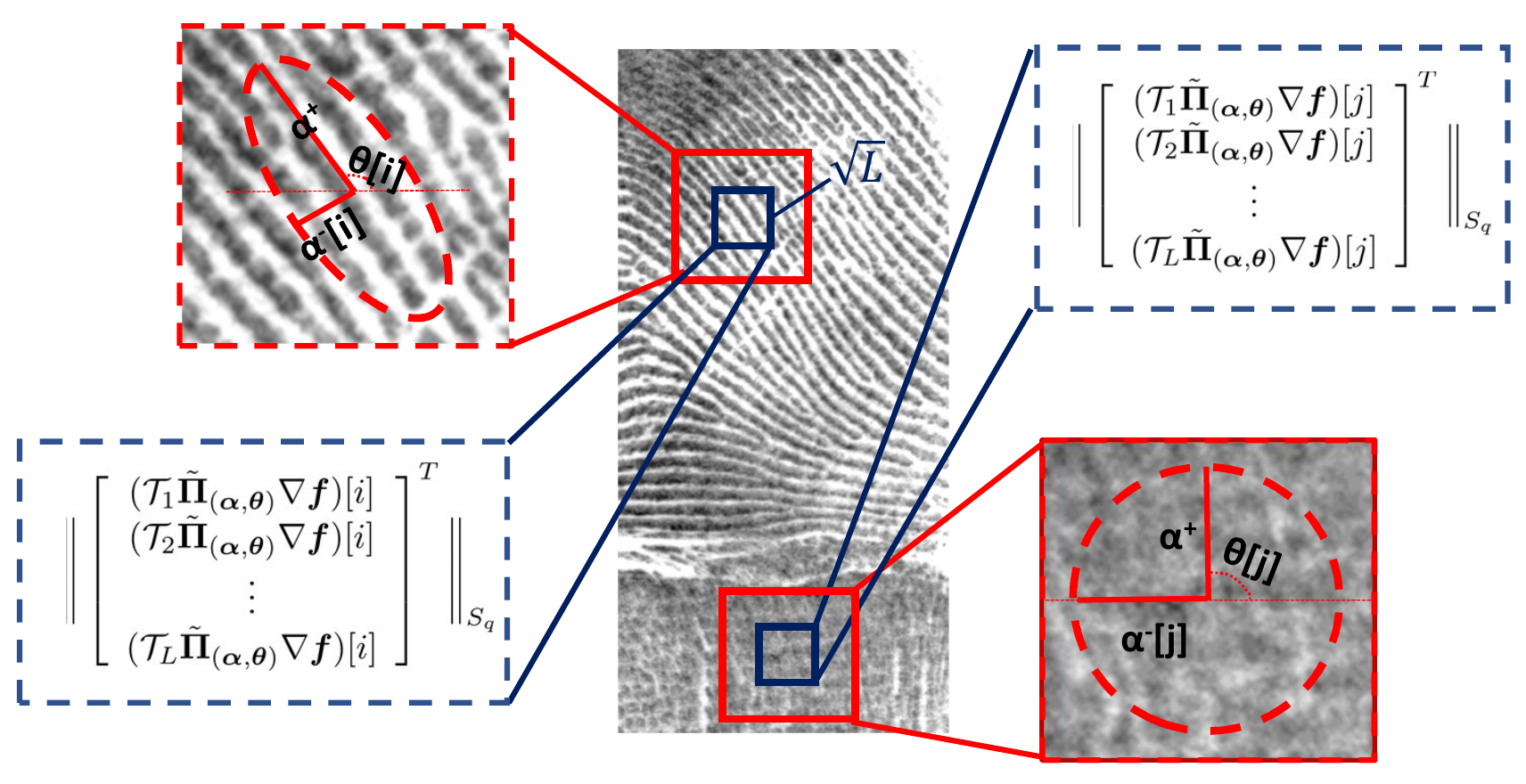}
  \caption{A rough illustration of the DPE procedure vs. ADSTV denoiser.}
 \label{fig:DPEvsADSTV}       
    \end{figure}

In Figure \ref{fig:DPE}, we present colormaps taken from the different stages of our DPE procedure applied to the examplary \textit{Zebra} image from Berkeley Segmentation Dataset (BSD) \cite{amfm_pami2011}, contaminated with the noise level $\sigma_\eta = 0.15$. Figure \ref{fig:DPE} (a) shows the coherence field extracted from the luminance image by using $\bar{\sigma} = \sqrt{11}$. Thanks to the neighborhood-awareness of the structure tensor, the directions of the small-scale structures are reliably distinguished without the need of pre-smoothing. Figure \ref{fig:DPE} (b) demonstrates the TV regularized coherence field after the application of Eq. \eqref{eq:coherenceTV}. One can see that the image points within highly directional $11 \times 11$ neighborhood respond with very high coherence values (e.g., the bodies of the zebras), while the response attenuates as the point gets closer to a nondirectional region (e.g. the contours of the bodies). This gradual decrease prevents the affects of the sharp transitions from the anisotropic behaviour to the isotropic one. Due to the uncertainities and the larger-scale directional structures that couldn't be extracted, the background is all suppressed. In Figure \ref{fig:DPE} (c), we show the colormap of the final $\boldsymbol{\alpha}^-$, obtained after applying one more sequence of Eq. \eqref{eq:coherence}, Eq. \eqref{eq:coherenceTV}, Eq. \eqref{eq:scaleSpaceCoherence}, and Eq. \eqref{eq:scaleSpaceCoherence2} at the scale with $\sigma_2$, and finally Eq. \eqref{eq:alphaFinal}. Additional to the result obtained at the scale $\sigma_1$, one can see the inclusion of some directional structures at the background encoded in orangish colors. Eventually, Figure \ref{fig:DPE} (d) shows the final distribution of the orientations, i.e., $\boldsymbol{\theta} = \hat{\boldsymbol{\theta}}$. Furthermore, in Figure \ref{fig:DPEvsADSTV}, we roughly illustrate the idea behind our two stage framework. The directional descriptors are visualized as an ellipse, whose major and minor axes respectively corresponds to the $\alpha^+$ and $\boldsymbol{\alpha}^-[i]$ at a point $i$, and its orientation coincides with the orientation to be penalized.

\subsection{Numerical Optimization}
According to the objective function given in Eq. \eqref{eq:energy}, the problem to be solved is:
\begin{equation}
\label{eq:DSTVproblem} 
\hat{\textbf{f}} = \underset{\textbf{f} \in \mathcal{C}}{\operatorname{argmin}} \ \frac{1}{2} \Vert \textbf{g} - \textbf{f}\Vert_2^2  + \tau \Vert \tilde{J}^{(\boldsymbol{\alpha},\boldsymbol{\theta})}_K \textbf{f} \Vert_{1,q} 
\end{equation}
where $\mathcal{C}$ is a set corresponding to an additional constraint ($\mathcal{C} = \mathbb{R}^{NC}$ in unconstrained case). The solution to this problem corresponds to the proximal map of $ADSTV(\textbf{f})$, i.e., $\text{prox}_{\tau ADSTV}(\textbf{g})$. 
\begin{comment}
\begin{equation}
\label{eq:proxADSTV} 
\text{prox}_{R}(\textbf{g}) = \underset{\textbf{f}\in \mathcal{C}}{\operatorname{arg min}} \ \frac{1}{2}\Vert \textbf{g} - \textbf{f} \Vert_2^2 + \tau \Vert \tilde{J}^{(\boldsymbol{\alpha},\boldsymbol{\theta})}_K \textbf{f} \Vert_{1,q} 
\end{equation}
\end{comment}
In \cite{demircan2019direction}, $\text{prox}_{\tau DSTV}(\textbf{g})$ was solved through the derivation of dual problem in detail. Having ADSTV instead of DSTV does not change the primal problem, hence the way that we derive dual will remain same. However, DSTV was only considering Frobenius norm ($q=2$). We define ADSTV in more general form, considering $q \geq 1$, as a direct extention of STV. In this respect, to come up with a concise derivation, using the fact that the dual of the norm $\Vert \cdot \Vert_{\mathcal{S}_q}$ is $\Vert \cdot \Vert_{\mathcal{S}_p}$, where $\frac{1}{q} + \frac{1}{p} = 1$, one can redefine ADSTV in terms of the support function of the form:
\begin{equation}
\label{eq:adstv_sup} 
ADSTV(\textbf{f}) = \underset{\Psi[i] \in B_{\mathcal{S}_p}}{\operatorname{sup}}  \langle \tilde{J}^{(\boldsymbol{\alpha},\boldsymbol{\theta})}_K \textbf{f}, \boldsymbol{\Psi} \rangle
\end{equation}
by introducing a variable $\boldsymbol{\Psi} \in \mathbb{R}^{N \times LC \times 2}$, and the set $B_{\mathcal{S}_p} = \lbrace \textbf{X} \in \mathbb{R}^{LC \times 2}: \Vert \textbf{X} \Vert_{S_p} \leq 1 \rbrace$. Here $\boldsymbol{\Psi}[i]$ refers to the $i$-th submatrix of $\boldsymbol{\Psi}$. This paves the way for rewriting the problem in Eq. \eqref{eq:DSTVproblem} as a minimax problem: 
\begin{equation}
\label{eq:minimax} 
\underset{\textbf{f} \in \mathcal{C}}{\operatorname{min}} \ \underset{\Psi[i] \in B_{\mathcal{S}_p}}{\operatorname{max}} \ \frac{1}{2}\Vert \textbf{g} - \textbf{f} \Vert_2^2 + \tau \langle \textbf{f}, \tilde{J}^{(\boldsymbol{\alpha},\boldsymbol{\theta})^*}_K \boldsymbol{\Psi} \rangle 
\end{equation}
%where $\mathcal{C}$ is nothing but a set that corresponds to an additional constraint on $f$ (e.g. box constraint), which is equal to $\mathbb{R}^{N}$ for unconstrained case.
where $\tilde{J}^{(\boldsymbol{\alpha},\boldsymbol{\theta})^*}_K$ arising after we leave $\textbf{f}$ alone in the second term denotes the adjoint. The adjoint of the patch-based Jacobian $J^*_K: \mathbb{R}^{N \times LC \times 2} \mapsto \mathbb{R}^{NC}$ was defined in \cite{lefkimmiatis2015structure} with its derivation:%A.3. Proof of Proposition 3.2.
\begin{equation}
\label{eq:patchJacobAdj} 
(J^*_K \textbf{X}) [t] = \sum_{l=1}^{L} -\text{div}(\mathcal{T}^*_{l}\textbf{X}[i,s])
\end{equation}
where $s = (c-1)L+l$ and $t=(c-1)N+n$ with $1 \leq n \leq N$ and $1 \leq c \leq C$.  $\mathcal{T}^*_{l}$ corresponds to the adjoint of $\mathcal{T}_{l}$, which scans the $\textbf{X}[i,s]$ in column-wise manner, where $\textbf{X}[i,s] \in \mathbb{R}^2$ is the $s$-th row of the $i$-th submatrix of an arbitrary $\textbf{X} \in \mathbb{R}^{N \times LC \times 2}$. Also, the operator $\text{div}$ is discrete divergence, applied by using backward differences, since the gradient is discretized using forward differences, as in \cite{lefkimmiatis2015structure}. From this point of view, in order to define our $\tilde{J}^{(\boldsymbol{\alpha},\boldsymbol{\theta})^*}_K: \mathbb{R}^{N \times LC \times 2} \mapsto \mathbb{R}^{NC}$, one should only change the divergence operator in Eq. \eqref{eq:patchJacobAdj} with the directional divergence operator $\widetilde{\text{div}}_{(\boldsymbol{\alpha}[i],\boldsymbol{\theta}[i])}$, i.e.,
\begin{equation}
\label{eq:directionalPatchJacobAdj} 
(\tilde{J}^{(\boldsymbol{\alpha},\boldsymbol{\theta})^*}_K \textbf{X}) [t] = \sum_{l=1}^{L} - \widetilde{\text{div}}_{(\boldsymbol{\alpha}[i],\boldsymbol{\theta}[i])}(\mathcal{T}^*_{l}\textbf{X}[i,s])
\end{equation}
whose definition is given below:
\begin{equation}
\label{eq:directionalDiv} 
\begin{split}
\widetilde{\text{div}}_{(\boldsymbol{\alpha}[i],\boldsymbol{\theta}[i])} (\cdot)  &\triangleq  \text{div} \ \big(\tilde{\boldsymbol{\Pi}}_{(\boldsymbol{\alpha}[i],\boldsymbol{\theta}[i])}^T (\cdot)\big) \\ &= \text{div} \ \big(\textbf{R}_{\boldsymbol{\theta}[i]} \tilde{\boldsymbol{\Lambda}}_{(\alpha^+,\boldsymbol{\alpha}^\textbf{-}[i])} (\cdot)\big)
\end{split}
\end{equation}

Let us call the objective function in Eq. \eqref{eq:minimax} as $\mathcal{L}(\textbf{f},\boldsymbol{\Psi})$. It is convex w.r.t. $\textbf{f}$ and concave w.r.t. $\boldsymbol{\Psi}$. By following the fast gradient projection (FGP) method \cite{beck2009fast}, which combines the dual approach introduced in \cite{chambolle2004algorithm} to solve TV-based denoising, and the fast iterative shrinkage-thresholding (FISTA) \cite{beck2009fista} to accelerate convergence, one can swap $\text{min}$ and $\text{max}$, i.e.,
\begin{equation}
\label{eq:saddle} 
\underset{\textbf{f} \in \mathcal{C}}{\operatorname{min}} \ \underset{\boldsymbol{\Psi}[i] \in B_{\mathcal{S}_p}}{\operatorname{max}} \ \mathcal{L}(\textbf{f},\Psi) = \mathcal{L}(\hat{\textbf{f}},\hat{\boldsymbol{\Psi}}) = \underset{\boldsymbol{\Psi}[i] \in B_{\mathcal{S}_p}}{\operatorname{max}} \ \underset{\textbf{f} \in \mathcal{C}}{\operatorname{min}} \ \mathcal{L}(\textbf{f},\boldsymbol{\Psi})
\end{equation}
since the common saddle point is not affected. Maximization of the dual problem $d(\boldsymbol{\Psi}) = {\operatorname{min}}_{\textbf{f} \in \mathcal{C}} \ \mathcal{L}(\textbf{f},\boldsymbol{\Psi})$ at the right-hand side is same with the minimization of the primal problem $P(\textbf{f}) = {\operatorname{max}}_{\boldsymbol{\Psi}[i] \in B_{\mathcal{S}_p}} \ \mathcal{L}(\textbf{f},\boldsymbol{\Psi})$ at the left-hand side. Therefore, finding the maximizer $\hat{\boldsymbol{\Psi}}$ of $d(\boldsymbol{\Psi})$ serves to find the minimizer $\hat{\textbf{f}}$ of $P(\boldsymbol{\Psi})$. One can rewrite $\hat{\textbf{f}}$ in terms of $\boldsymbol{\Psi}$, as:
 \begin{equation}
\label{eq:dual} 
\hat{\textbf{f}} = \underset{\textbf{f} \in \mathcal{C}} {\operatorname{argmin}}  \ \Vert \textbf{f} - (\textbf{g} - \tau \tilde{J}_K^{(\boldsymbol{\alpha},\boldsymbol{\theta})^*}{\boldsymbol{\Psi}}) \Vert^2_2 - M
\end{equation}
by expanding $\mathcal{L}(\textbf{f},\boldsymbol{\Psi})$ and collecting the constants under the term $M$. The solution to Eq. \eqref{eq:dual} is:
 \begin{equation}
\label{eq:dualgrad} 
\hat{\textbf{f}} = P_{\mathcal{C}}({\textbf{g} - \tau \tilde{J}_K^{(\boldsymbol{\alpha},\boldsymbol{\theta})^*}{\boldsymbol{\Psi}}})
\end{equation}
where $P_{\mathcal{C}}$ is the orthogonal projection onto the set $\mathcal{C}$. Then we proceed by plugging $\hat{\textbf{f}}$ in $\mathcal{L}(\textbf{f},\boldsymbol{\Psi})$ to get $d(\boldsymbol{\Psi}) = \mathcal{L}(\hat{\textbf{f}},\boldsymbol{\Psi})$, i.e., 
 \begin{equation}
\label{eq:dualProblem} 
d(\boldsymbol{\Psi}) = \frac{1}{2} \Vert \textbf{w} - P_\mathcal{C}(\textbf{w})\Vert^2_2 + \frac{1}{2} (\Vert \textbf{g} \Vert^2_2 - \Vert \textbf{w} \Vert^2_2)
\end{equation}
where $\textbf{w} = \textbf{g} - \tau \tilde{J}_K^{(\boldsymbol{\alpha},\boldsymbol{\theta})^*} \boldsymbol{\Psi}$. Dual problem is smooth with a well-defined gradient computed:
\begin{equation}
\label{eq:dualGradient} 
\nabla d(\boldsymbol{\Psi}) = \tau \tilde{J}_K^{(\boldsymbol{\alpha},\boldsymbol{\theta})} P_\mathcal{C}(\textbf{g} - \tau \tilde{J}_K^{(\boldsymbol{\alpha},\boldsymbol{\theta})^*}{\boldsymbol{\Psi}})
\end{equation} 
based on the derivation in Lemma 4.1 of \cite{beck2009fast}. Therefore, in order to find the maximizer $\hat{\boldsymbol{\Psi}}$ of $d(\boldsymbol{\Psi})$, one can employ the projected gradients \cite{chambolle2004algorithm}. It performs decoupled sequences of gradient descent/ascent and projection onto a set. For the gradient ascent in our case, an appropriate step size that ensures the convergence needs to be selected. For the gradient in Eq. \eqref{eq:dualGradient}, a ${LC \times 2}$ matrix filled by a constant step size $1/\textbf{L}[i]$ can be used for each image point, where $\textbf{L} \in \mathbb{R}^{N \times LC \times 2}$ is a multidimensional array of Lipschitz constants each of which having the upper bound $\textbf{L}[i] \leq 8\sqrt{2}\tau^2 \big(({{\alpha}^{+}})^2+({\boldsymbol{\alpha}^{-}[i])}^2\big)$. One can see Appendix A for the derivation.

When it comes to the projection, as an efficient way, the authors of \cite{lefkimmiatis2015structure} reduced the projection of each $\boldsymbol{\Psi}[i]$ onto $B_{\mathcal{S}_p}$ to the projection of the singular values onto the $\ell_p$ unit ball $B_p = \lbrace \boldsymbol{\sigma} \in \mathbb{R}^2_{+}:\Vert \boldsymbol{\sigma} \Vert_p \leq 1\rbrace$. Therefore, for the SVD $\boldsymbol{\Psi}[i] = \textbf{U} \boldsymbol{\Sigma} \textbf{V}^T$ and $\boldsymbol{\Sigma} = \text{diag}(\boldsymbol{\sigma})$, where $\boldsymbol{\sigma} = [\sigma_1,\sigma_2]$, the projection was defined as:
\begin{equation}
\label{eq:projection} 
P_{B_{\mathcal{S}_p}}(\boldsymbol{\Psi}[i]) = \boldsymbol{\Psi}[i] \boldsymbol{V} \boldsymbol{\Sigma}^+ \boldsymbol{\Sigma}_\textbf{p} \textbf{V}^T
\end{equation}
where $\boldsymbol{\Sigma}^+$ is pseudoinverse of $\boldsymbol{\Sigma}$ and $\boldsymbol{\Sigma}_p = \text{diag}(\boldsymbol{\sigma}_p)$ for $\boldsymbol{\sigma}_p$ are the projected singular values of $\boldsymbol{\Sigma}$. The readers are referred to \cite{lefkimmiatis2015structure} for the details about the derivation and the realization. 
%\begin{equation}
%\label{eq:projection} 
%P_{B_{\mathcal{S}_p}}(\boldsymbol{\Psi}[i]) = \frac{\boldsymbol{\Psi}[i]}{\text{max}(1,\Vert \boldsymbol{\Psi}[i] \Vert_{\mathcal{S}_p})}
%\end{equation}
As a consequence, the overall algorithm is shown in Algorithm 1. 

\begin{algorithm}[t!]
\caption{Algorithm for ADSTV-based denoising}\label{euclid}
\begin{spacing}{1.1}
\begin{algorithmic}
\small
\STATE \textbf{INPUT:} $\textbf{g}$, ${\alpha}^+ > 1$, $\tau > 0$, $p > 1$, $P_\mathcal{C}$ 
\STATE \textbf{INIT:} ${\boldsymbol{\Psi}}^{(0)} = \textbf{0} \in \mathbb{R}^{N \times LC \times 2}$, \quad $t^{(1)} = 1$, \quad $i=1$ %$\epsilon \in (0,min\lbrace 1,L(d)\rbrace ), \quad \gamma \in [\epsilon, 2L(d) - \epsilon], \quad \tau \in [\epsilon,1]$ 
\STATE $(\boldsymbol{\theta}, \boldsymbol{\alpha^-}) \leftarrow $ \textbf{DPE}$(\textbf{g}, \alpha^+)$ [Section~\ref{DPE}]
\STATE $\textbf{L} \leftarrow {8\sqrt{2}\tau\big(({\alpha^+})^2 +({\boldsymbol{\alpha}^-})^{\circ 2}\big)}$ 
%\STATE $\bar{v} \leftarrow \mathcal{F}^{-1}((\widehat{H}^{*} \widehat{H})^{-1} \widehat{H}^{*} \widehat{{g}})$
%\STATE $\bar{g}_U^{(k=0)} \leftarrow (\mathcal{P}\bar{v})_U$
\WHILE{stopping criterion is not satisfied}
\STATE $\textbf{z} \leftarrow P_\mathcal{C}(\textbf{g} - \tau \tilde{J}_K^{(\boldsymbol{\alpha},\boldsymbol{\theta})^*}{\boldsymbol{\Psi}^{(i-1)}})$
\STATE ${\boldsymbol{\Psi}}^{(i)} \leftarrow P_{B_{S_p}}\Big ({\boldsymbol{\Psi}}^{(i-1)} + \textbf{L}^{\circ-1} \odot \big ( \tilde{J}_K^{(\boldsymbol{\alpha},\boldsymbol{\theta})} \textbf{z} \big )\Big )$  
\STATE $({\boldsymbol{\Psi}}^{(i+1)}, t^{(i+1)}) \leftarrow$ \textbf{FISTA} ($t^{(i)}, {\boldsymbol{\Psi}}^{(i)}, {\boldsymbol{\Psi}}^{(i-1)}$)
\STATE $i \leftarrow i + 1$
\ENDWHILE
\STATE return $P_C(\textbf{g}^{(k)} - \tau \tilde{J}_K^{(\boldsymbol{\alpha},\boldsymbol{\theta})^*} {\boldsymbol{\Psi}}^{(j)})$
%\STATE $\bar{g}_U^{(k)} \leftarrow  (\mathcal{F}^{-1}(\widehat{\mathcal{P}}\widehat{H}\widehat{f}^{(k)}))_U$
\\
\hrulefill
\STATE \textbf{\small {function}} \textbf{FISTA} ($t, \textbf{f}^{\text{(cur)}}, \textbf{f}^{\text{(prev)}}$) \cite{beck2009fista}
\STATE \quad $t^{\text{(next)}} \leftarrow (1+\sqrt{1+4t^2})\big /2$
\STATE \quad $\textbf{f}^{\text{(next)}} \leftarrow \textbf{f}^{\text{(cur)}} + (\frac{t - 1}{t^{\text{(next)}}})(\textbf{f}^{\text{(cur)}} - \textbf{f}^{\text{(prev)}})$
\STATE \quad return $(\textbf{f}^{\text{(next)}}, t^{\text{(next)}})$
\STATE \textbf{\small {end}}
\end{algorithmic}
\end{spacing}
\end{algorithm}

\section{Experiments}
In this section, we assess qualitative and quantitative performances of the proposed variational denoising framework. We compare it with the other related analysis-based regularizers: TV (as a baseline), EADTV (as a representative of previous attempts to make DTV \cite{bayram2012directional} adaptive), STV \cite{lefkimmiatis2015structure} (as a predecessor of ADSTV), and NCDR \cite{liu2018image} (as a recent analysis-based regularization scheme depending on a new paradigm). We don't include TGV \cite{bredies2010total} in the competing algorithms since, STV's superiority over it has already been demonstrated in \cite{lefkimmiatis2015structure}. Owing to the fact that TV, EADTV, and NCDR are only applicable to the scalar-valued images, they are only involved in the grayscale environment. Thus, the experiments on the vector-valued images merely compare STV and DSTV regularizers. The source codes of STV  \footnote{\url{https://github.com/cig-skoltech/Structure_Tensor_Total_Variation}} and NCDR \footnote{\url{https://github.com/HUST-Tan/NCDR}} that we use were made publicly available by the authors on GitHub. Our ADSTV is implemented on top of STV, while TV and EADTV are written from scratch. All methods were written in MATLAB by only making use of the CPU. The runtimes were computed on a computer equipped with Intel Core Processor i7-7500U (2.70-GHz) with 16 GB of memory.

\begin{figure*}[t!]
\includegraphics[width=\textwidth]{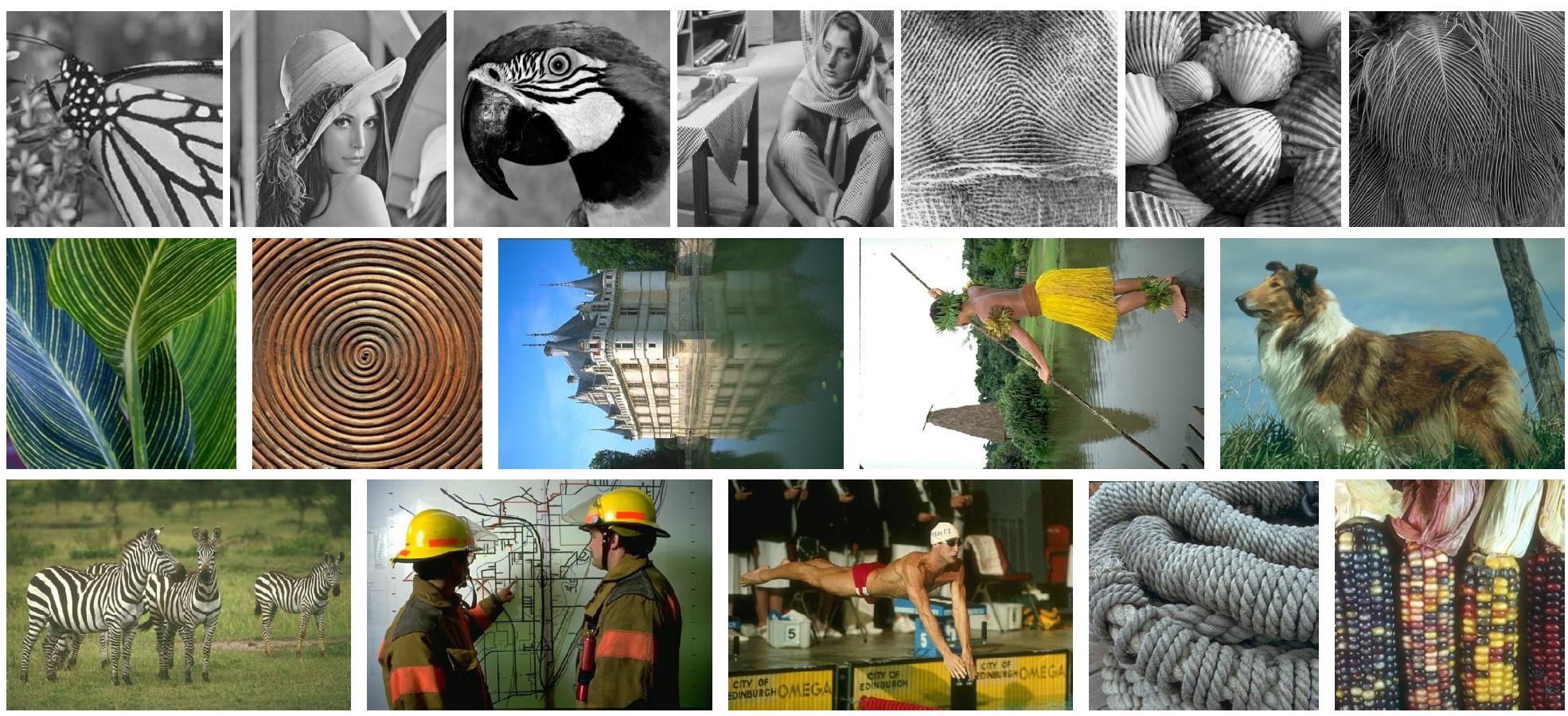}
  \caption{{Thumbnails of the grayscale and color images used in the experiments.} From left to right and top to bottom: \textit{Monarch, Lena, Parrot, Barbara, Fingerprint, Shells, Feather, Plant, Spiral, Chateau, Indigenous, Dog, Zebra, Workers, Swimmer, Rope}, and \textit{Corns}.}
  \label{fig:images} 
\end{figure*}

 To assess the quantitative performances of the methods, we use peak signal-to-noise ratio (PSNR) in dB and structural similarity index (SSIM). The experiments are performed on the images shown in Figure \ref{fig:images}. The first row shows the thumbnails of seven classical grayscale test images of sizes $256 \times 256$ (\textit{Monarch}, \textit{Shells}, and \textit{Feather}) and $512 \times 512$ (the others). In the other two rows, we have RGB color images. \textit{Plant, Spiral, Rope} and \textit{Corns} of sizes $256 \times 256$ are more textural with different directional characteristics. \textit{Rope} and \textit{Corns} are taken from the popular Berkeley Segmentation Dataset (BSD) \cite{amfm_pami2011}, while the others are public domain images. The rest of the color images are natural images taken from BSD. They have either no or limited amount of directional parts. Note that, the intensities of all test images have been normalized to the range $[0, 1]$.

The algorithms under consideration are all aiming to minimize Eq. \eqref{eq:energy}. Therefore, the critical regularization parameter $\tau$ is in common, and should be chosen precisely. For a fair comparison, we fine-tuned $\tau$ such that it leads to the best PSNR in all experiments. For STV, ADSTV, and NCDR priors; one should also choose the convolution kernel. We used a Gaussian kernel of support $3 \times 3$ for all three algorithms. The standard deviations ($\sigma$) were set to $0.5$ for STV and DSTV, while $\sigma = 1$ was used for NCDR, as suggested in the algorithms' respective papers: \cite{lefkimmiatis2015structure} and \cite{liu2018image}. Both STV and ADSTV use the nuclear norm of the patch-based Jacobian, since it was chosen as the best performing norm. Similar to the convolution kernel, all the other parameters of NCDR were kept same as they were used in \cite{liu2018image}. The EADTV's $\alpha$ is fine-tuned by changing it from 2 (since it reduces to TV when $\alpha = 1$) to 30 with 1-unit intervals. This is also the case for the ADSTV's $\alpha^+$, unless otherwise stated. DSTV requires an additional convolution kernel too, to be used by the preprocessor DPE. We fixed it as a Gaussian kernel with variance and support $\bar{\sigma}^2 = 7$ for $256 \times 256$ images, while $\bar{\sigma}^2 = 11$ for $963 \times 481$ natural images from BSD, and $\bar{\sigma}^2 = 15$ for $512 \times 512$ images. For the optimization of all TV-based algorithms, the stopping criterion was either having a relative normed difference between two successive estimates that reaches a threshold of $10^{-5}$, or fulfilling  a number of iterations, equals to 100 in our case. When it comes to NCDR, since it uses gradient descent to solve its non-convex optimization problem, it is quite hard to ensure that whether a reported result is due to a local or a global minimum. The number of iterations for NCDR was experimentally set to 500.

% Table generated by Excel2LaTeX from sheet 'grayscale-paperv2'
\begin{table*}[t!]
  \centering
  \caption{PSNR/SSIM Comparison of TV, EADTV, STV, NCDR, and ADSTV on grayscale images. Last column shows the average runtimes.}
    \resizebox{\textwidth}{!}{\begin{tabular}{c|c|p{4.215em}p{4.215em}p{4.215em}p{4.215em}p{4.215em}p{4.215em}p{4.215em}|p{4.215em}|l}
    \hline
    \hline
          &       & \multicolumn{1}{l}{Monarch} & \multicolumn{1}{l}{Lena} & \multicolumn{1}{l}{Parrot} & \multicolumn{1}{l}{Barbara} & \multicolumn{1}{l}{Fingerp.} & \multicolumn{1}{l}{Shell} & \multicolumn{1}{l|}{Feather} & \multicolumn{1}{l|}{\textbf{Avg.}} &
          \multicolumn{1}{l}{\textbf{t (sec.)}}\\
    \hline
    \hline
    \parbox[t]{2mm}{\multirow{5}[2]{*}{\rotatebox[origin=c]{90}{$\sigma_\eta = 0.05$}}} & TV    & 31.54/0.91 & 32.88/0.87 & 33.17/0.88 & 29.48/0.84 & 29.14/0.93 & 28.69/0.89 & 26.79/0.93 & 30.24/0.89 & 1.51 \\
          & EADTV & 31.91/0.92 & 33.09/0.87 & 33.16/0.88 & 29.59/0.83 & 29.73/0.94 & 28.81/0.89 & 26.86/0.93 & 30.45/0.89 & 1.82\\
          & STV   & 32.32/\textbf{0.93} & 33.59/0.88 & 33.56/0.90 & 30.19/0.88 & \textbf{30.11/0.94} & 29.01/0.89 & 27.16/0.94 & 30.85/0.91 & 17.41\\
          & NCDR  & \textbf{32.66/0.93} & \textbf{34.26/0.89} & 33.40/0.90 & \textbf{31.44/0.89} & 30.08/\textbf{0.94} & 29.24/0.89 & \textbf{28.47}/0.94 & \textbf{31.36}/0.91 & 276.98\\
          & ADSTV  & 32.61/\textbf{0.93}  & 34.03/\textbf{0.89}  & \textbf{33.77/0.91 } & 31.09/\textbf{0.89} & 30.09/\textbf{0.94}  & \textbf{29.36/0.90 } & 27.89/\textbf{0.95} & 31.25/\textbf{0.92} & 38.24\\
    \hline
    \parbox[t]{2mm}{\multirow{5}[2]{*}{\rotatebox[origin=c]{90}{$\sigma_\eta = 0.1$}}} & TV    & 27.88/0.85 & 29.86/0.81 & 29.74/0.82 & 25.57/0.71 & 25.33/0.86 & 24.94/0.77 & 21.95/0.81 & 26.47/0.80 & 2.08\\
          & EADTV & 28.31/0.84 & 30.07/0.81 & 29.81/0.81 & 25.51/0.70 & 25.83/0.87 & 24.45/0.76 & 21.65/0.81 & 26.52/0.80 & 2.77\\
          & STV   & 28.65/0.87 & 30.54/0.83 & 30.22/0.84 & 26.45/0.77 & 26.21/0.87 & 25.34/0.79 & 22.40/0.83 & 27.12/0.83 & 16.92\\
          & NCDR  & \textbf{29.29/0.88} & \textbf{31.36/0.84} & 30.31/\textbf{0.85} & \textbf{27.56}/0.80 & \textbf{26.54/0.88} & 25.74/0.79 & \textbf{24.32}/0.84 & \textbf{27.88}/0.84 & 256.77 \\
          & ADSTV  & 29.07/\textbf{0.88} & 31.15/\textbf{0.84} & \textbf{30.47/0.85} & {27.34}/\textbf{0.81} & 26.48/\textbf{0.88} & \textbf{25.83/0.81} & 23.60/\textbf{0.86} & 27.70/\textbf{0.85} & 34.59\\
    \hline
    \parbox[t]{2mm}{\multirow{5}[2]{*}{\rotatebox[origin=c]{90}{$\sigma_\eta = 0.15$}}} & TV    & 25.84/0.80 & 28.23/0.77 & 27.91/0.77 & 23.88/0.64 & 23.30/0.79 & 23.06/0.68 & 19.58/0.67 & 24.54/0.73 & 2.28\\
          & EADTV & 26.24/0.80 & 28.38/0.77 & 28.02/0.77 & 23.78/0.63 & 24.22/0.82 & 23.22/0.69 & 19.03/0.67 & 24.70/0.74 & 3.07\\
          & STV   & 26.58/0.82 & 28.85/0.79 & 28.43/0.79 & 24.47/0.69 & 24.20/0.81 & 23.48/0.71 & 20.04/0.70 & 25.15/0.76 & 16.77\\
          & NCDR  & \textbf{27.34/0.84} & \textbf{29.61}/0.80 & 28.69/0.80 & \textbf{25.47}/0.72 & \textbf{24.64/0.82} & {24.00}/0.70 & \textbf{22.00}/0.72 & \textbf{25.96}/0.77 & 270.70\\
          & ADSTV  & 27.02/\textbf{0.84} & 29.45/\textbf{0.81} & \textbf{28.73/0.81} & 25.40/\textbf{0.74} & 24.46/0.81 & \textbf{24.04}/\textbf{0.73} & 21.30/\textbf{0.76} & 25.76/\textbf{0.79} & 31.63 \\
    \hline
    \parbox[t]{2mm}{\multirow{5}[2]{*}{\rotatebox[origin=c]{90}{$\sigma_\eta = 0.2$}}} & TV    & 24.43/0.75 & 27.14/0.74 & 26.68/0.73 & 23.04/0.61 & 21.94/0.72 & 21.87/0.61 & 18.21/0.54 & 23.33/0.67 & 2.78\\
          & EADTV & 24.78/0.76 & 27.22/0.74 & 26.77/0.73 & 22.96/0.60 & 22.80/0.77 & 22.00/0.62 & 18.12/0.54 & 23.50/0.68 & 2.58 \\
          & STV   & 25.13/0.78 & 27.69/0.76 & 27.22/0.76 & 23.45/0.65 & 22.68/0.76 & 22.27/0.64 & 18.65/0.58 & 23.87/0.70 & 16.21\\
          & NCDR  & \textbf{25.92/0.80} & \textbf{28.35/0.78} & 27.56/0.77 & \textbf{24.21}/0.66 & \textbf{23.28/0.77} &\textbf{22.81}/0.63 & \textbf{20.35}/0.60 & \textbf{24.64}/0.72 & 269.34\\
          & ADSTV  & 25.61/\textbf{0.80} & 28.29/\textbf{0.78} & \textbf{27.59/0.78} & 23.98/\textbf{0.67} & 23.12/\textbf{0.77} & 22.79/\textbf{0.66} & 19.66/\textbf{0.65} & 24.43/\textbf{0.73} & 36.77\\
    \hline
    \parbox[t]{2mm}{\multirow{5}[2]{*}{\rotatebox[origin=c]{90}{$\sigma_\eta = 0.25$}}} & TV    & 23.35/0.72 & 26.33/0.71 & 25.75/0.71 & 22.52/0.58 & 20.93/0.67 & 21.02/0.55 & 17.38/0.42 & 22.47/0.62 & 2.76\\
          & EADTV & 23.64/0.72 & 26.35/0.72 & 25.82/0.71 & 22.48/0.58 & 21.68/0.72 & 21.13/0.57 & 17.23/0.41 & 22.62/0.63 & 2.82\\
          & STV   & 24.01/0.75 & 26.82/0.74 & 26.30/0.74 & 22.90/0.61 & 21.58/0.70 & 21.40/0.59 & 17.75/0.47 & 22.97/0.66 & 16.60\\
          & NCDR  & \textbf{24.80/0.77} & \textbf{27.37}/0.75 & 26.65/0.75 & \textbf{23.33}/0.62 & \textbf{22.16/0.71} & {21.88}/0.57 & \textbf{19.11}/0.48 & \textbf{23.61}/0.66 & 274.48\\
          & ADSTV  & 24.53/\textbf{0.77} & \textbf{27.37}/\textbf{0.76} & \textbf{26.78/0.76} & 23.22/\textbf{0.63} & 22.11/\textbf{0.71} & \textbf{21.90/0.61} & 18.54/\textbf{0.54} & 23.49/\textbf{0.68} & 36.70 \\
    \hline
    \hline
    \end{tabular}}%
  \label{tab:grayscale}%
\end{table*}%

\begin{figure*}[t!]
\includegraphics[width=\textwidth]{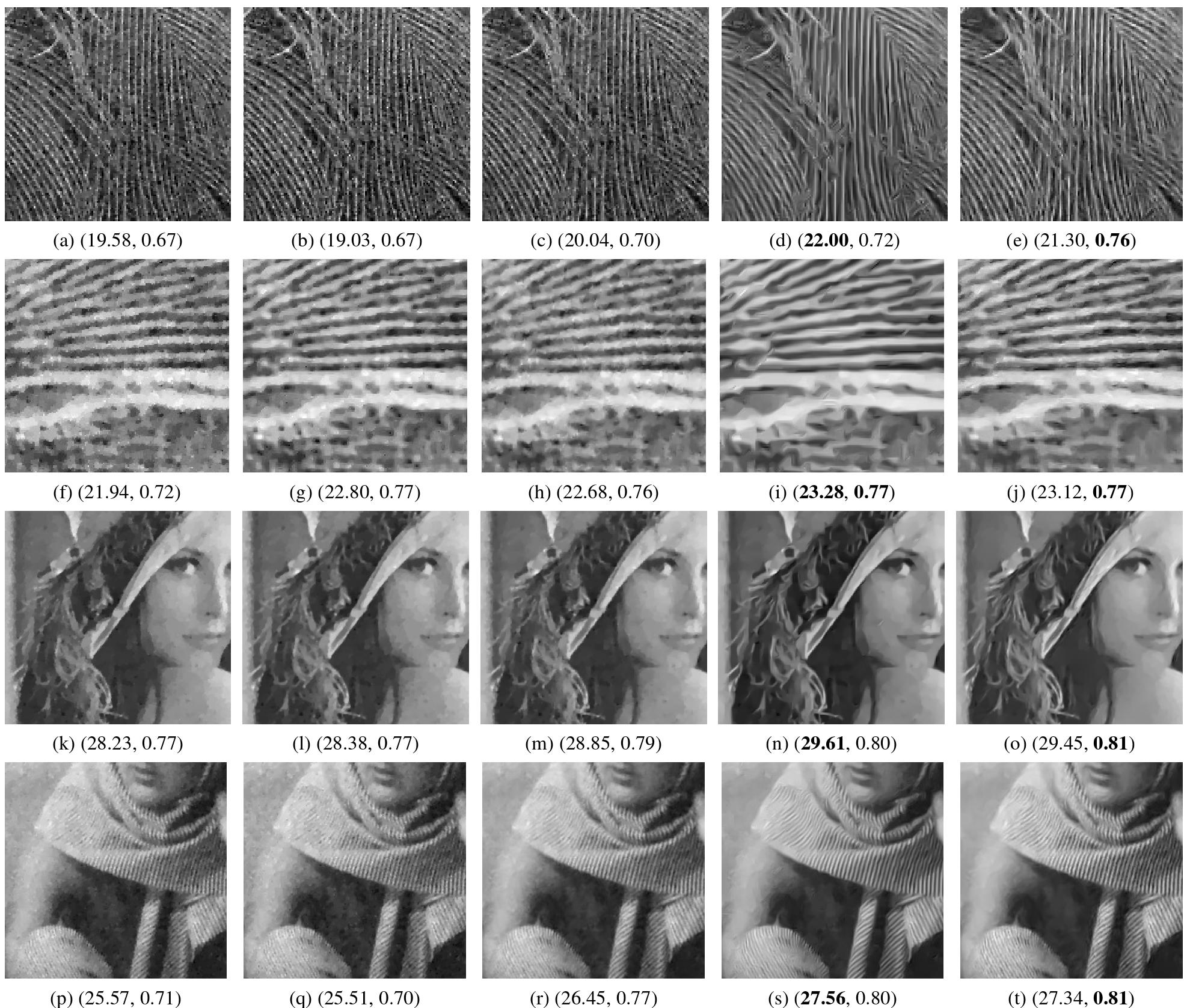}
\caption{The detail patches showing grayscale image denoising results. The noisy \textit{Barbara} ($\sigma_\eta = 0.1$), \textit{Feather, Lena} ($\sigma_\eta = 0.15$), and \textit{Fingerprint} ($\sigma_\eta = 0.2$) images are restored by using TV (col-1), EADTV (col-2), STV (col-3), NCDR (col-4), and ADSTV (col-5) regularizers. The quantity pairs shown at the bottom of each image are corresponding to the (PSNR, SSIM) values.}
 \label{fig:grayscaledenoising} 
\end{figure*}

We consider additive i.i.d. Gaussian noise with five noise levels of $\sigma_\eta = \lbrace$ 0.05, 0.1, 0.15, 0.2, 0.25$\rbrace $. In Table \ref{tab:grayscale}, we report PSNR and SSIM measures obtained by using TV, EADTV, STV, NCDR, and ADSTV priors applied to the grayscale images. From the results, we observe that the EADTV does not offer significant improvements over TV in the images with less directional patches such as \textit{Lena} and \textit{Parrot}. But, it even performs worse than TV for \textit{Barbara} and \textit{Feather} images, in which highly directional parts present, as the noise level increases. This is due to the fact that, those directional parts involve very high frequency components. Thus, even small deviations in the direction estimation extremely affects the result. In almost all of the experiments, EADTV preferred to use $\alpha = 2$ (except \textit{Fingerprint} for which larger $\alpha$ values are selected), and this is also due to the need of compensating the sheer amount of mistakes made by its direction estimator. STV produced significantly better results when compared to TV and EADTV for all cases and images, except \textit{Fingerprint}, where for all the noise levels $\sigma_\eta > 0.1$, EADTV outperformed STV. When it comes to NCDR, it yielded superior PSNR results to all competing algorithms including ADSTV, except the experiments on \textit{Parrot} and \textit{Shell}. However, as one may see in the last column, these results are obtained at the cost of computational load. The average runtimes of NCDR at all noise levels are pretty high. On the other hand, when the quality is measured by SSIM, our ADSTV systematically outperformed NCDR (and the others). Figure \ref{fig:grayscaledenoising} is provided for the visual judgement. It can be inferred that NCDR smooths the structural regions better than ADSTV at the risk of loss of details (See (d)-(e), (i)-(j), and the scarf of \textit{Barbara} in (s)-(t)). NCDR also causes artifacts in the junctions more apperant than those of ADSTV (e.g., (d)-(e)). Also the ADSTV is better at smoothing flat region as one can see on \textit{Lena}'s face in (n)-(o). These may clarify why the SSIM values obtained by ADSTV are higher.

\begin{table*}[t!]%htbp
  \centering
  \caption{PSNR/SSIM Comparison of STV and ADSTV on color image denoising. Last row shows the average runtimes.}
    \resizebox{\textwidth}{!}{\begin{tabular}{l|ll|ll|ll|ll|ll}
    \hline
    \hline
      & \multicolumn{2}{c|}{$\sigma_\eta = 0.05$} & \multicolumn{2}{c|}{$\sigma_\eta = 0.1$} & \multicolumn{2}{c|}{$\sigma_\eta = 0.15$} & \multicolumn{2}{c|}{$\sigma_\eta = 0.2$} & \multicolumn{2}{c}{$\sigma_\eta = 0.25$} \\
    \hline
   & \multicolumn{1}{c}{STV} & \multicolumn{1}{c|}{ADSTV} & \multicolumn{1}{c}{STV} & \multicolumn{1}{c|}{ADSTV} & \multicolumn{1}{c}{STV} & \multicolumn{1}{c|}{ADSTV} & \multicolumn{1}{c}{STV} & \multicolumn{1}{c|}{ADSTV} & \multicolumn{1}{c}{STV} & \multicolumn{1}{c}{ADSTV} \\
    \hline
    \hline
    Plant & \multicolumn{1}{p{4.215em}}{29.23/0.96} & \multicolumn{1}{p{4.215em}|}{\textbf{30.32/0.97}} & \multicolumn{1}{p{4.215em}}{25.27/0.90} & \multicolumn{1}{p{4.215em}|}{\textbf{26.63/0.93}} & \multicolumn{1}{p{4.215em}}{23.28/0.85} & \multicolumn{1}{p{4.215em}|}{\textbf{24.60/0.89}} & \multicolumn{1}{p{4.215em}}{22.05/0.80} & \multicolumn{1}{p{4.215em}|}{\textbf{23.24/0.85}} & \multicolumn{1}{p{4.215em}}{21.21/0.76} & \multicolumn{1}{p{4.215em}}{\textbf{22.27/0.81}} \\
    \hline
    Spiral & \multicolumn{1}{p{4.215em}}{27.82/0.97} & \multicolumn{1}{p{4.215em}|}{\textbf{28.39/0.98}} & \multicolumn{1}{p{4.215em}}{24.02/0.93} & \multicolumn{1}{p{4.215em}|}{\textbf{24.76/0.94}} & \multicolumn{1}{p{4.215em}}{22.04/0.89} & \multicolumn{1}{p{4.215em}|}{\textbf{23.06/0.91}} & \multicolumn{1}{p{4.215em}}{20.73/0.86} & \multicolumn{1}{p{4.215em}|}{\textbf{21.95/0.89}} & \multicolumn{1}{p{4.215em}}{19.74/0.83} & \multicolumn{1}{p{4.215em}}{\textbf{21.08/0.87}} \\
    \hline
    Chateau & \multicolumn{1}{p{4.215em}}{32.43/0.95} & \multicolumn{1}{p{4.215em}|}{\textbf{32.74/0.96}} & \multicolumn{1}{p{4.215em}}{28.83/0.91} & \multicolumn{1}{p{4.215em}|}{\textbf{29.39/0.92}} & \multicolumn{1}{p{4.215em}}{26.93/0.87} & \multicolumn{1}{p{4.215em}|}{\textbf{27.63/0.89}} & \multicolumn{1}{p{4.215em}}{25.75/0.85} & \multicolumn{1}{p{4.215em}|}{\textbf{26.46/0.87}} & \multicolumn{1}{p{4.215em}}{24.88/0.83} & \multicolumn{1}{p{4.215em}}{\textbf{25.51/0.85}} \\
    \hline
    Indigenous & \multicolumn{1}{p{4.215em}}{32.70/0.95} & \multicolumn{1}{p{4.215em}|}{\textbf{33.04/0.96}} & \multicolumn{1}{p{4.215em}}{29.33/0.90} & \multicolumn{1}{p{4.215em}|}{\textbf{29.74/0.92}} & \multicolumn{1}{p{4.215em}}{27.71/0.88} & \multicolumn{1}{p{4.215em}|}{\textbf{28.07/0.89}} & \multicolumn{1}{p{4.215em}}{26.70/0.85} & \multicolumn{1}{p{4.215em}|}{\textbf{26.99/0.87}} & \multicolumn{1}{p{4.215em}}{25.99/0.84} & \multicolumn{1}{p{4.215em}}{\textbf{26.25/0.85}} \\
    \hline
    Dog   & \multicolumn{1}{p{4.215em}}{32.19/0.95} & \multicolumn{1}{p{4.215em}|}{\textbf{32.53/0.96}} & \multicolumn{1}{p{4.215em}}{29.00/0.90} & \multicolumn{1}{p{4.215em}|}{\textbf{29.40/0.91}} & \multicolumn{1}{p{4.215em}}{27.44/0.87} & \multicolumn{1}{p{4.215em}|}{\textbf{27.80/0.88}} & \multicolumn{1}{p{4.215em}}{26.46/0.84} & \multicolumn{1}{p{4.215em}|}{\textbf{26.76/0.85}} & \multicolumn{1}{p{4.215em}}{25.78/0.82} & \multicolumn{1}{p{4.215em}}{\textbf{26.05/0.83}} \\
    \hline
    Zebra & \multicolumn{1}{p{4.215em}}{30.70/\textbf{0.97}} & \multicolumn{1}{p{4.215em}|}{\textbf{30.87/0.97}} & \multicolumn{1}{p{4.215em}}{26.86/0.93} & \multicolumn{1}{p{4.215em}|}{\textbf{27.46/0.94}} & \multicolumn{1}{p{4.215em}}{24.84/0.90} & \multicolumn{1}{p{4.215em}|}{\textbf{25.58/0.91}} & \multicolumn{1}{p{4.215em}}{23.46/0.86} & \multicolumn{1}{p{4.215em}|}{\textbf{24.26/0.89}} & \multicolumn{1}{p{4.215em}}{22.47/0.84} & \multicolumn{1}{p{4.215em}}{\textbf{23.34/0.87}} \\
    \hline
    Workers & \multicolumn{1}{p{4.215em}}{31.92/0.95} & \multicolumn{1}{p{4.215em}|}{\textbf{32.41/0.96}} & \multicolumn{1}{p{4.215em}}{28.06/0.90} & \multicolumn{1}{p{4.215em}|}{\textbf{28.77/0.92}} & \multicolumn{1}{p{4.215em}}{25.96/0.85} & \multicolumn{1}{p{4.215em}|}{\textbf{26.77/0.89}} & \multicolumn{1}{p{4.215em}}{24.57/0.81} & \multicolumn{1}{p{4.215em}|}{\textbf{25.39/0.85}} & \multicolumn{1}{p{4.215em}}{23.56/0.77} & \multicolumn{1}{p{4.215em}}{\textbf{24.37/0.82}} \\
    \hline
    Swimmer & \multicolumn{1}{p{4.215em}}{33.63/0.97} & \multicolumn{1}{p{4.215em}|}{\textbf{33.92/0.98}} & \multicolumn{1}{p{4.215em}}{30.17/\textbf{0.95}} & \multicolumn{1}{p{4.215em}|}{\textbf{30.57/0.95}} & \multicolumn{1}{p{4.215em}}{28.29/0.92} & \multicolumn{1}{p{4.215em}|}{\textbf{28.73/0.93}} & \multicolumn{1}{p{4.215em}}{27.03/0.90} & \multicolumn{1}{p{4.215em}|}{\textbf{27.43/0.92}} & \multicolumn{1}{p{4.215em}}{26.09/0.88} & \multicolumn{1}{p{4.215em}}{\textbf{26.50/0.90}} \\
    \hline
    Rope  & \multicolumn{1}{p{4.215em}}{28.37/\textbf{0.91}} & \multicolumn{1}{p{4.215em}|}{\textbf{28.53/0.91}} & \multicolumn{1}{p{4.215em}}{24.68/0.80} & \multicolumn{1}{p{4.215em}|}{\textbf{24.96/0.81}} & \multicolumn{1}{p{4.215em}}{22.87/0.70} & \multicolumn{1}{p{4.215em}|}{\textbf{23.18/0.72}} & \multicolumn{1}{p{4.215em}}{21.73/0.61} & \multicolumn{1}{p{4.215em}|}{\textbf{22.06/0.63}} & \multicolumn{1}{p{4.215em}}{20.93/0.55} & \multicolumn{1}{p{4.215em}}{\textbf{21.23/0.58}} \\
    \hline
    Corns & \multicolumn{1}{p{4.215em}}{29.74/\textbf{0.95}} & \multicolumn{1}{p{4.215em}|}{\textbf{29.90/0.95}} & \multicolumn{1}{p{4.215em}}{25.94/0.89} & \multicolumn{1}{p{4.215em}|}{\textbf{26.16/0.90}} & \multicolumn{1}{p{4.215em}}{23.92/0.83} & \multicolumn{1}{p{4.215em}|}{\textbf{24.06/0.84}} & \multicolumn{1}{p{4.215em}}{22.59/0.78} & \multicolumn{1}{p{4.215em}|}{\textbf{22.74/0.79}} & \multicolumn{1}{p{4.215em}}{21.63/\textbf{0.74}} & \multicolumn{1}{p{4.215em}}{\textbf{21.77/0.74}} \\
    \hline
    \hline
    \textbf{Avg. } & 30.87/0.95 & \textbf{31.27/0.96} & 27.22/0.90 & \textbf{27.78/0.91} & 25.33/0.86 & \textbf{25.95/0.88} & 24.11/0.82 & \textbf{24.73/0.84} & 23.23/0.79 & \textbf{23.84/0.81} \\
    \hline
    \textbf{t (sec.) } & \multicolumn{1}{c}{20.62} & \multicolumn{1}{c|}{40.18} & \multicolumn{1}{c}{21.32} & \multicolumn{1}{c|}{33.81} & \multicolumn{1}{c}{18.97} & \multicolumn{1}{c|}{33.39} & \multicolumn{1}{c}{21.12} & \multicolumn{1}{c|}{36.27} & \multicolumn{1}{c}{22.50} & \multicolumn{1}{c}{35.54} \\
    \hline
    \hline
    \end{tabular}}%
  \label{tab:color}%
\end{table*}%

\begin{figure*}[t!]
\centering
\includegraphics[width=17cm]{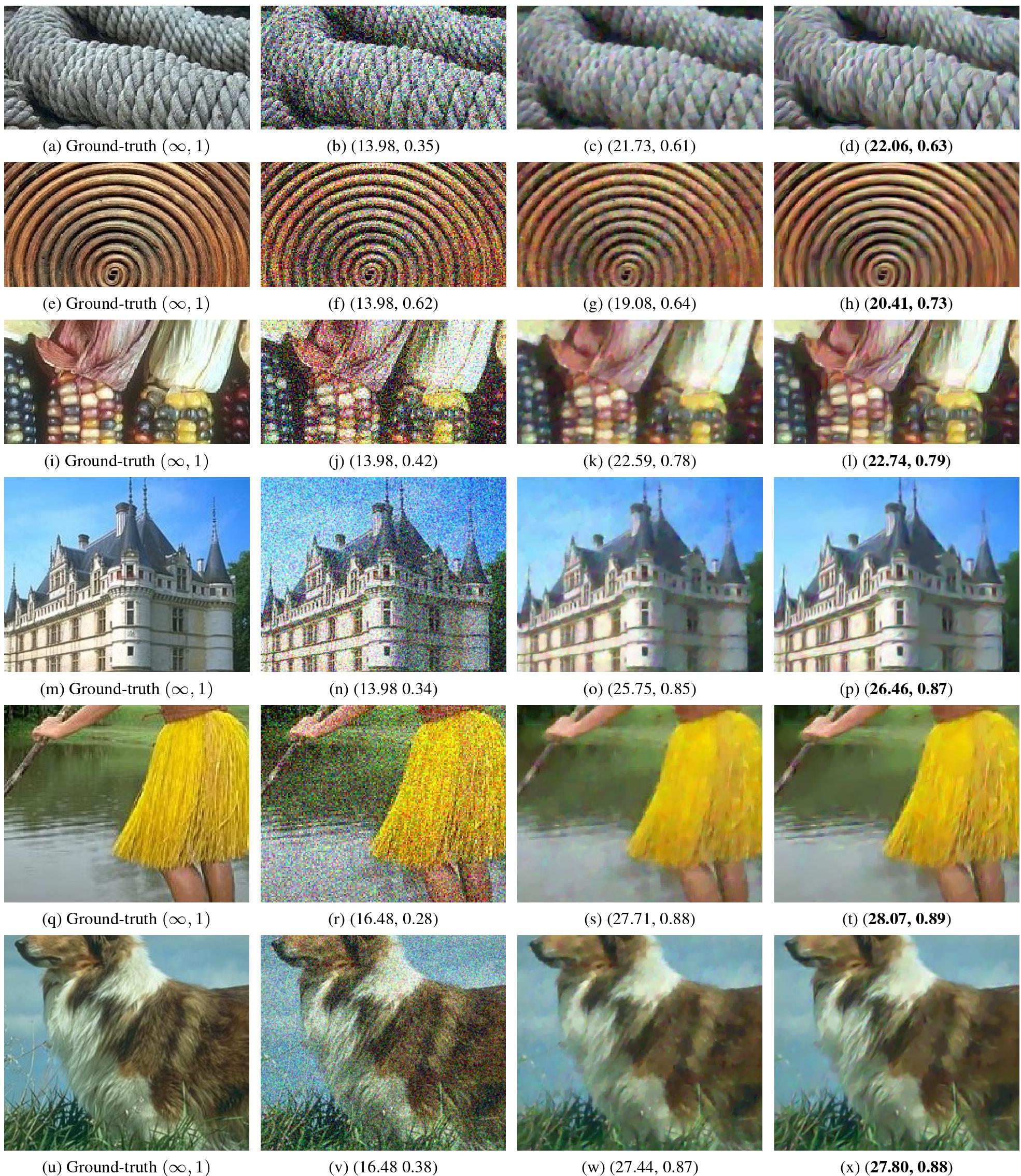}
\caption{The detail patches showing color image denoising results. The noisy \textit{Indigenous, Dog} ($\sigma_\eta = 0.15$), and \textit{Rope, Spiral, Corns, Chateau, Indigenous} ($\sigma_\eta = 0.2$) images (col-2) are restored by using STV (col-3) and ADSTV (col-4) regularizers. The quantity pairs shown at the bottom of each image are corresponding to the (PSNR, SSIM) values.}
 \label{fig:colordenoising} 
 \end{figure*}

Table \ref{tab:color} on the other side, shows PSNR/SSIM values obtained by the application of STV and ADSTV to vector-valued images. According to the results, ADSTV systematically outperformed STV in terms of PSNR and SSIM measures, even in the images that do not exhibit directional dominance. However, in terms of the runtimes that are reported in the last row, ADSTV seems to bring almost twice as much load to the computation. Again in Figure \ref{fig:colordenoising}, we demonstrate exemplary detail patches cropped from the original images (first column), noisy versions (second column), and the restored versions by STV (third column) and ADSTV (last column). The results obtained by STV have oil-painting-like artifacts, whereas this effect is far less visible in ADSTV's results. With the incorporation of the directional information, the edges became more apparent, (see \textit{Rope} and \textit{Chateau}), more smoothing towards the desired direction (see  \textit{Spiral}), and the gaps between closely parallel lines could better be distinguished (e.g. the feather in \textit{Dog} image, the details of the leaves in \textit{Corns} image, and skirt in \textit{Indigenous} image). 

\begin{figure*}[t!]
\centering
\newcommand{\mywidth}{8cm}
\captionsetup[subfigure]{labelformat=empty}
\subfloat[(a) $\sigma_\eta = 0.15$]{\includegraphics[width=\mywidth]{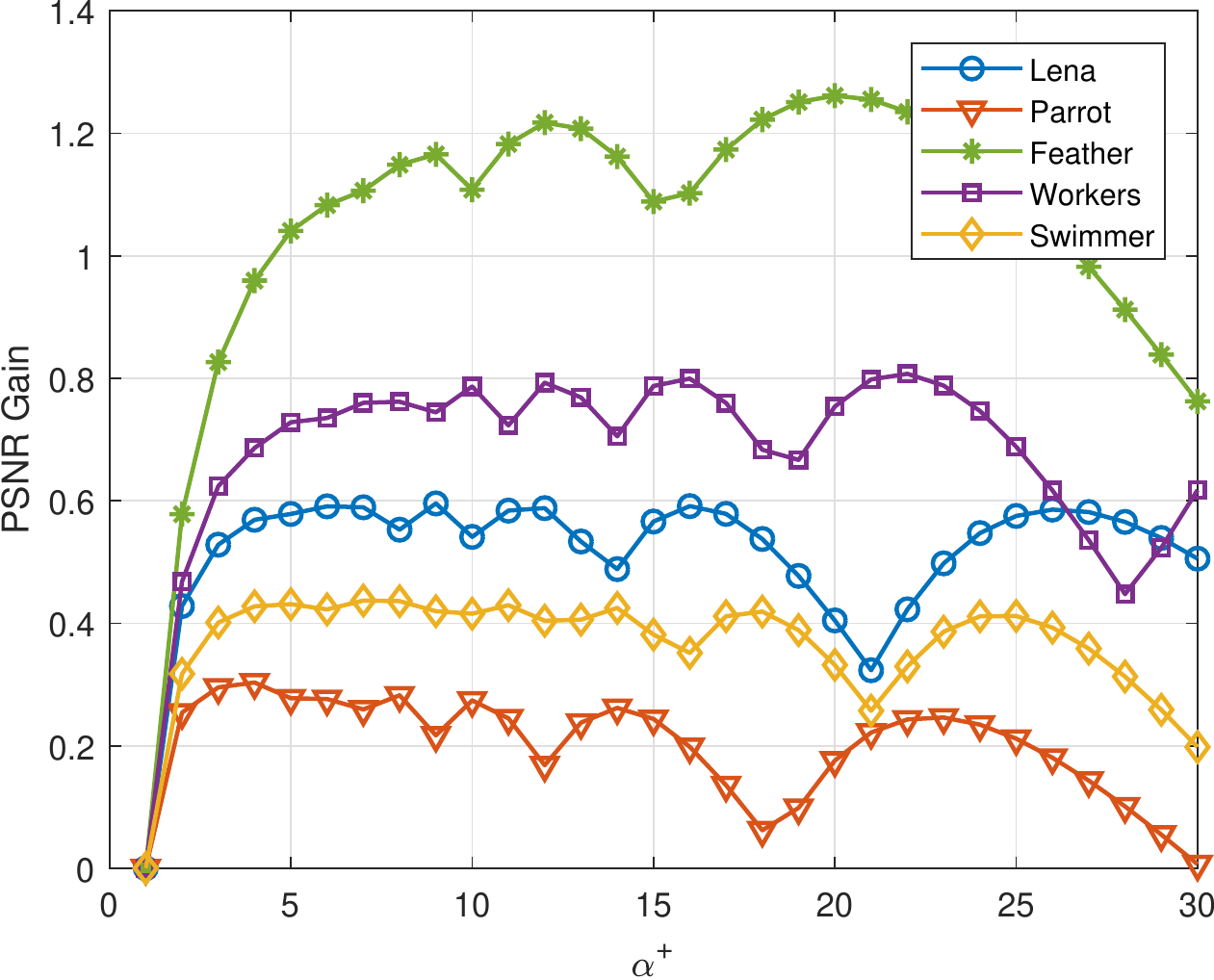}}\hspace{1cm}
\subfloat[(b) \textit{Barbara}]{\includegraphics[width=\mywidth]{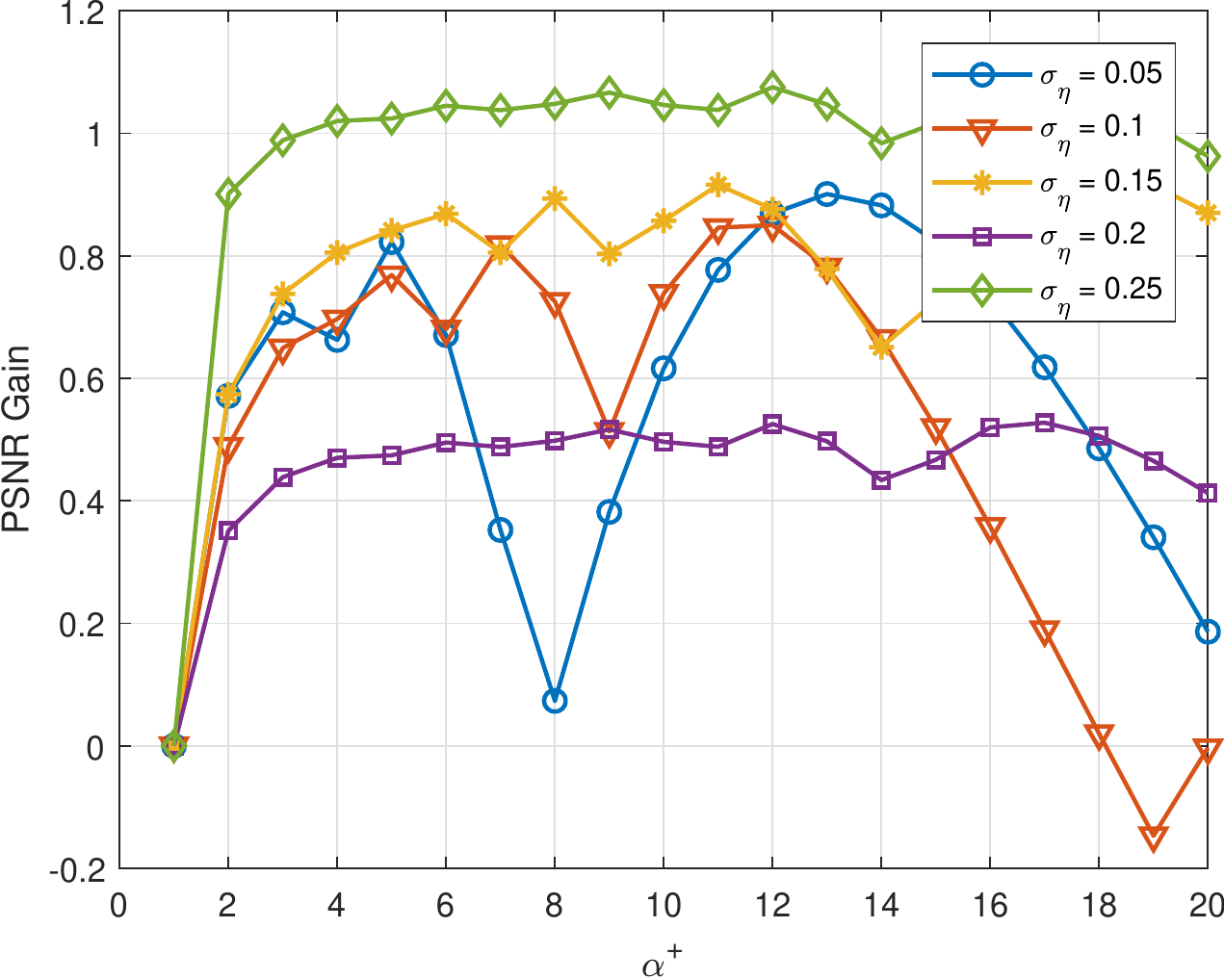}}
\caption{PSNR gain over STV with respect to $\alpha^+$.}
\label{fig:alphavsPSNRGain} 
\end{figure*}
      
One of the ADSTV's disadvantages is introducing an extra free parameter $\alpha^+$ that needs to be tuned. In Figure \ref{fig:alphavsPSNRGain}, we present two graphs to show how sensitive the gain over STV by using our method with respect to $\alpha^+$ is. Figure \ref{fig:alphavsPSNRGain} (a) serves the purpose by testing varying $\alpha^+$ values on five different grayscale and color images at the noise level $\sigma_\eta = 0.15$, while Figure \ref{fig:alphavsPSNRGain} (b) examines them on \textit{Barbara} image at five different noise levels. As one can infer from the results, even for the incorrect selections of $\alpha^+$, the gain has almost always been above zero, which shows the robustness of our approach. Yet a relevant research question here might be if it is possible to calculate $\alpha^+$ automatically, as a function of an estimated noise level by using variance estimators and coherence. 
%which both have locally dominant directions. The presence of the noise causes unreliable estimation of the directions.  DSTV yields the best performance in all noise levels and in all images. By inspecting the results, one can deduce that the DSTV produces the performance of DSTV prominently increases when the subjected image has directional patches inside.

\section{Discussion}
In the previous section, we compared the proposed denoising framework with TV \cite{rudin1992nonlinear}, EADTV \cite{bayram2012directional, zhang2013edge}, STV \cite{lefkimmiatis2015structure}, and NCDR \cite{liu2018image} based denoising models. STV was holding the state-of-the-art records among the other TV-related local analysis-based priors. NCDR came up with a new paradigm regarding the sparsity of the corner points, in 2018. Although the NCDR puts plausible results out, its non-convexity and heuristic nature make it disadvantageous. Moreover, it's only applicable to the scalar-valued images. The qualitative results and SSIM measurements show the effectiveness of our ADSTV over NCDR in terms of restoration quality, while NCDR results in better PSNR. But when we consider the computational performances, NCDR machinery is around 7 times slower than ADSTV. We believe that our framework provides a good balance between the restoration quality and the computational performance.    

On the other hand, as mentioned in Sect. \ref{sec:intro} and Sect. \ref{subsec:DTV}, a closely related paper \cite{pang2020image} was published as prepress (to be published in 2020) at the moment that we were studying on this paper. That work couldn't be involved in our experiments but, due to the common experiment performed on the same \textit{Lena} image, we think it is worth talking over it here. The authors reported that the restored versions of \textit{Lena} contaminated by Gaussian noise with $\sigma = \lbrace 0.05, 0.1, 0.15 \rbrace$ has the signal to noise ratio (SNR) around $17.5, 15$ and, $13$,  respectively; whereas the corresponding results of ADSTV are $19.77, 16.84$, and $15.07$. We think this difference is due to the incorporation of structure tensor.

Finally, in order not to finish discussion without referring to today's learning based approaches on restoration, we want to touch upon briefly how close our results to those obtained by the popular feed-forward denoising convolutional neural networks (DnCNN) \cite{zhang2017beyond}. On the same \textit{Parrot}, \textit{Lena}, and \textit{Barbara} images of the range $[0, 255]$, the authors of DnCNN experimented three noise levels $\lbrace 15, 25, 50 \rbrace$ and reported the PSNR scores. When we perform the same experiments on the \textit{Parrot} image, we observed that the results of ADSTV were almost 1 dB better than DnCNN. However on \textit{Lena} and \textit{Barbara} images it was not the case, and DnCNN performed around 1 dB and 1.5 dB better than ADSTV, respectively, as might be expected. Similar to the classification tasks, data-driven methods seems to predominate the model-driven ones in restoration tasks. They can themselves capture the underlying geometry within the data. Moreover, they can run very fast once trained. However, beside those advantages, they also have limitations such as the lack of the suitable ground truth images that would be used for training purpose, difficulty of incorporating prior domain knowledge, and the strong dependency to the imaging parameters (e.g. noise level) of the training data. To tackle with these limitations, new efforts try to involve the variational models in the learning-based methods \cite{liu2018image, lefkimmiatis2018universal, ulyanov2018deep}. Also as remarked in \cite{mccann2017convolutional}, the learning-based approaches comprise the variational approaches
as special cases. These facts and the non-negligible drawbacks motivate the researchers to produce in both fields with the hope that their achievements might further be combined. %That keeps the image restoration based on the variational models as an open problem.

\section{Conclusion}
In this study, we've described a two-stage variational framework for denoising scalar- and vector-valued images. First stage is the preprocessor that estimates some directional parameters that will guide to the second stage, and the second stage is the one that performs denoising. Our preprocessor attempts to circumvent the chicken-and-egg problem of capturing local geometry from a noisy image to be used while denoise it, by employing the eigenvalues and the orthogonal eigenvectors of the structure tensor, and making assumptions on the piecewise-constancy of the underlying directional parameters. When it comes to the denoising stage, it encodes the problem as a convex optimization problem that involves the proposed adaptive direction-guided structure tensor total variation (ADSTV) as regularizer. Our ADSTV is designed by extending the structure tensor total variation (STV), so that it works biased in a certain direction varying at each point in the image domain. This approach results in improved restoration quality that has been verified by extensive experiments. We compared our method with the related denoising models involving various analysis-based priors.

We believe that the ADSTV regularizer can be used in a large spectrum of inverse imaging problems, even though the effectiveness of our preprocessor may change from problem to problem. Furthermore, there might be appropriate application fields where the source image that is used to estimate the directional parameters differs from the target image to be restored (e.g., video restoration where the source and target frames are different, pansharpening where the chromatic image can be used as a source for the parameters). We plan to pursue this issue as future work.

%should be explored further.

%%%%%%%%%%%%%%%%%%%%%%%%%%%%%%%%%%%%%%%%%%%%%%
%%                                          %%
%% Backmatter begins here                   %%
%%                                          %%
%%%%%%%%%%%%%%%%%%%%%%%%%%%%%%%%%%%%%%%%%%%%%%

\section*{Appendix A}
In order to find an upper bound for Lipschitz constants, we follow the derivation in \cite{beck2009fast} and adapt it to our formulation.
\small
\begin{equation}
\label{eq:back_stepsize}
\begin{split}
\Vert \nabla d(\textbf{u}) - \nabla d(\textbf{v}) \Vert &=  \tau \Vert \tilde{J}_K^{(\boldsymbol{\alpha},\boldsymbol{\theta})} P_\mathcal{C}(\textbf{g} - \tau \tilde{J}_K^{(\boldsymbol{\alpha},\boldsymbol{\theta})^*}{\textbf{u}}) - \tilde{J}_K^{(\boldsymbol{\alpha},\boldsymbol{\theta})} P_\mathcal{C}(\textbf{g} - \tau \tilde{J}_K^{(\boldsymbol{\alpha},\boldsymbol{\theta})^*}{\textbf{v}}))\Vert\\
& \leq \tau \Vert \tilde{J}_K^{(\boldsymbol{\alpha},\boldsymbol{\theta})} \Vert \Vert P_\mathcal{C}(\textbf{g} - \tau \tilde{J}_K^{(\boldsymbol{\alpha},\boldsymbol{\theta})^*}{\textbf{u}}) - P_\mathcal{C}(\textbf{g} - \tau \tilde{J}_K^{(\boldsymbol{\alpha},\boldsymbol{\theta})^*}{\textbf{v}}))\Vert\\
& \leq \tau \Vert \tilde{J}_K^{(\boldsymbol{\alpha},\boldsymbol{\theta})} \Vert \Vert \tau \tilde{J}_K^{(\boldsymbol{\alpha},\boldsymbol{\theta})^*} (\textbf{u}-\textbf{v}) \Vert\\
& \leq \tau^2 \Vert \tilde{\Pi}_{(\boldsymbol{\alpha},\boldsymbol{\theta})} \Vert^2 \Vert J_K \Vert^2 \Vert \textbf{u}-\textbf{v} \Vert\\
& \leq \tau^2 \Vert \tilde{\Pi}_{(\boldsymbol{\alpha},\boldsymbol{\theta})} \Vert^2 \Vert \nabla \Vert^2 \Vert T\Vert^2 \Vert \textbf{u}-\textbf{v} \Vert\\
\end{split}
\end{equation}
\normalsize
%$\tilde{\Pi}_{(\boldsymbol{\alpha}[i],\boldsymbol{\theta}[i])} = \tilde{\boldsymbol{\Lambda}}_{(\alpha^+,\boldsymbol{\alpha}^\textbf{-}[i])} \textbf{R}_{\boldsymbol{-\theta}[i]}$
where $T = \sum_{l=1}^L (\mathcal{T}_{l}^* \mathcal{T}_{l})$. Knowing from \cite{beck2009fast} that  $\Vert \nabla \Vert^2 \leq 8$, from \cite{lefkimmiatis2015structure} that $\Vert T \Vert^2 \leq \sqrt{2}$, and further showing that $\Vert \tilde{\Pi}_{(\boldsymbol{\alpha},\boldsymbol{\theta})} \Vert ^2 \leq {(\alpha^+)}^2 + ({\boldsymbol{\alpha}^-)}^{\circ 2}$, we come up with the spatially varying Lipschitz constants ${\textbf{L}}[i] \leq 8\sqrt{2} \tau^2 ({(\alpha^+)}^2 + {(\boldsymbol{\alpha}^-[i])}^2)$. 



%% References



%% New version of the num-names style
\section*{Acknowledgements}
The authors would like to thank the anonymous reviewers for their valuable comments and suggestions. This work was supported by The Scientific and Technological Research Council of Turkey (TUBITAK) under 115R285.

\bibliographystyle{elsarticle-num-names}
\bibliography{sample}






% \end{thebibliography}

\end{document}